\RequirePackage{rotating}
\documentclass[fleqn,usenatbib]{mnras}

\usepackage{amssymb}	
\usepackage{multicol}        
\usepackage{bm}		
\usepackage{pdflscape}	
\usepackage{graphicx}
\usepackage{subfigure}
\usepackage{epstopdf}
\usepackage{amsmath}
\usepackage[mathscr]{eucal}
\usepackage{longtable}
\usepackage{wrapfig}
\usepackage{multirow}
\usepackage{hyperref}

\newcommand{\Msun}{M_{\odot}}
\newcommand{\Lsun}{L_{\odot}}

\newcommand{\kms}{\mbox{${\rm km~s^{-1}}$}}
\newcommand{\kmsMpc}{\mbox{km s$^{-1}$ Mpc$^{-1}$}}

\newcommand{\Pc}{\mathcal{P}}
\newcommand{\Dc}{\mathcal{D}}
\newcommand{\Mc}{\mathcal{M}}
\newcommand{\Lc}{\mathcal{L}}
\newcommand{\hi}{H{\sc\,i} }

\usepackage{newtxtext,newtxmath}

\usepackage[T1]{fontenc}

\DeclareRobustCommand{\VAN}[3]{#2}
\let\VANthebibliography\thebibliography
\def\thebibliography{\DeclareRobustCommand{\VAN}[3]{##3}\VANthebibliography}

\title{Cosmicflows-4: The Baryonic Tully-Fisher Relation Providing $\sim 10,000$ Distances}

\author[E. Kourkchi et al.]{
Ehsan Kourkchi,$^{1}$\thanks{E-mail: ehsan20@hawaii.edu}
R. Brent Tully,$^{1}$\thanks{E-mail: tully@ifa.hawaii.edu}
H\'el\`ene M. Courtois,
Alexandra Dupuy,$^{2}$
Daniel Guinet$^{2}$
\\
$^{1}$Institute for Astronomy, University of Hawaii, 2680 Woodlawn Drive, Honolulu, HI 96822, USA\\
$^{2}$University of Lyon, UCB Lyon 1, IUF, CNRS/IN2P3,  IP2I Lyon, UMR5822, F-69622 Villeurbanne. France
}

\date{Accepted XXX. Received YYY; in original form ZZZ}

\pubyear{2021}

\begin{document}
\label{firstpage}
\pagerange{\pageref{firstpage}--\pageref{lastpage}}
\maketitle

\begin{abstract}
The interstellar gas in spiral galaxies can constitute a significant fraction of the baryon mass and it has been demonstrated that the sum of stellar and gas components correlates well with the kinematic signature of the total mass content, the widths of \hi line profiles.  The correlation of  baryonic mass with \hi line widths is used here to obtain distances for 9984 galaxies extending to $\sim 0.05c$.  The sample is \hi flux limited and a correction is required to account for an \hi selection bias.  The absolute scale is established by 64 galaxies with known distances from studies of Cepheid variables and/or the magnitudes of stars at the tip of the red giant branch. The calibration of the baryonic relationship results in a determination of the Hubble constant of $H_0=75.5\pm2.5$~\kmsMpc.  The error estimate is statistical.  This material will be combined with contributions from other methodologies in a subsequent article where systematic uncertainties will be investigated.

\end{abstract}


\section{Introduction}

The empirical coupling between the luminosities of spiral galaxies and their rotation rates \citep{1977A&A....54..661T} (TFR) has an evident zeroth order explanation: more intrinsically luminous galaxies tend to be more massive and more massive galaxies rotate faster under equilibrium conditions.  \citet{2000ApJ...533L..99M} recognized that optical or infrared light represents only the stellar component of mass, that contributions from interstellar gas could be important, and the star and gas together give representation of the total baryonic mass of a galaxy.  
The correlation between the baryonic mass and the rotation rate of a galaxy has become known as the baryonic Tully-Fisher relation (BTFR). The mainstream opinion is that a third component besides stars and gas, dark matter, generally dominates the mass budget, creating a challenge to theorists of galaxy formation as to why the correlation between baryons alone and kinematics should be so tight \citep{2005MNRAS.363.1299O, 2007MNRAS.374.1479G, 2011ApJ...742...16T, 2011ApJ...742...76G, 2014MNRAS.437.1750M, 2020MNRAS.498.3687G}.  
Indeed, most of the interest in the BTFR by both theorists and observers has been on the constraints it provides to our understanding of galaxy formation and evolution.  
By contrast, the limited concern of this paper, one that has only received slight attention \citep{2020AJ....160...71S}, is not to fully understand the physics of the BTFR but just to use the empirically tight BTFR to measure galaxy distances.

This study follows an investigation of the strictly stellar TFR at {\it Sloan Digital Sky Survey} (SDSS) optical and {\it Wide-field Infrared Space Explorer} (WISE) passbands \citep{2020ApJ...902..145K}.  
The galaxy samples used then and now, the photometry, the inclinations, the reddening assumptions, and the \hi linewidths, are the same.
The additional component, neutral hydrogen fluxes, is gathered both from observations within our collaboration and from those by others as discussed in $\S$\ref{sec:gas}.


Until now, we had favored the optical or infrared luminosity correlation with rotation velocity, the TFR, for the measurement of distances because of its two parameter simplicity (although multi-band couplings have been exploited to reduce scatter by \citet{2019ApJ...884...82K, 2020ApJ...896....3K, 2020ApJ...902..145K}).  It was our impression that gas contributions would only be important with relatively faint galaxies that drop out at systemic velocities in the range 
 $2,000-15,000$~\kms\ that are of particular interest for BTFR distance applications. 
However, as will be shown, that impression is misplaced.  Gas contributions can be significant over a wide range of spiral luminosities.  The impetus for the present study, though, was finding a trend in the Hubble parameter, $H_j = cz_j /d_j$, with apparent magnitudes in the \citet{2020ApJ...902..145K} study, where $d_j$ are distances determined by the photometric correlations with rotation.  
The mean Hubble parameter $<H_j>$ decreases to lower values in the faintest apparent magnitude bins, even though $<H_j>$ holds satisfactorily constant in redshift bins.

The trend could be traced to curvature in the TFR.
Curvature has been suspected from early days \citep{1982ApJS...50..241A,1993ApJ...418..626P,2007MNRAS.381.1463N,2011ApJ...742...16T}.  In our own work, we have treated curvature as a bending at the bright end and applied bias corrections to distance estimates derived from simulations of a bent TFR with assumed errors
\citep{2014ApJ...792..129N, 2020ApJ...902..145K}. Our procedure has been to determine distances to galaxies in the field using a fiducial relation formed from a calibration sample drawn from galaxies in a modest number of well studied clusters \citep{2020ApJ...896....3K}.  This cluster calibration sample, though it has grown to contain 600 galaxies in 20 clusters, only hints at the curvature problem.  Plausibly it could be taken to be only a bright end effect.  However, we now see that the curvature must be viewed as a continuous bending with intrinsic luminosity.

\begin{figure}
\centering
\includegraphics[width=0.95\linewidth]{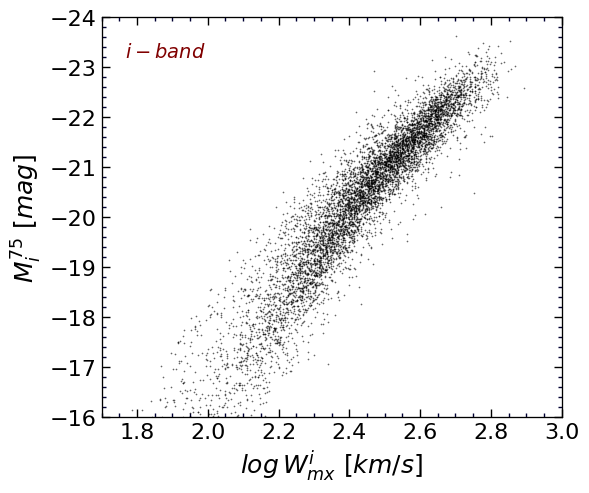}
\caption{Correlation between SDSS $i$ magnitudes and linewidths with absolute magnitudes calculated assuming $H_0=75$~\kmsMpc.  Extreme outliers (1.4\% of sample) have been rejected.
\label{fig:tf75}}
\end{figure}

This reality is demonstrated in Figure~\ref{fig:tf75}.  The $i$-band TFR for our entire SDSS sample (with a $3.5\sigma$ clip) is plotted at Hubble flow distances assuming 
the mean observed velocity of a galaxy's group and $H_0 = 75$~\kmsMpc.
Curvature in the TFR is evident.  Procedures could be explored to compensate for the curvature in the measurement of distances.  However, the addition of a gas mass term preferentially boosts  the fainter galaxies
that tend to have large gas fractions \citep{2006ApJ...653..240G} and has been demonstrated to produce a relatively power law relation between baryonic mass and rotation rate \citep{2005ApJ...632..859M, 2016ApJ...816L..14L, 2019MNRAS.484.3267L}.  Here, we exploit the usefulness of the BTFR to determine distances to $\sim10^4$ galaxies.

\subsection{Outline of the Discussion}

The BTFR invokes a correlation between the combined contributions of the stars and gas in a galaxy with rotation rate, a measure of the total mass of the system.  The observational extension on the TFR is the incorporation of \hi fluxes.  The sources of this information are discussed in \S \ref{sec:gas}.
Issues regarding the inference of stellar mass from observed luminosities are discussed in \S \ref{sec:stellar_mass}.  Luminosities are acquired at both optical and infrared bands and consistency is required in mass inferences across bands. The formulations derived in this section are tentative.  Refined relations are given by the Bayesian analysis discussed in \S \ref{sec:optimizeOP} and \S \ref{sec:optimizeIR}.  The dominant source of bias in the BTFR procedure arises from a faintness detection limit on \hi fluxes.  The nature of this bias is discussed in \S \ref{sec:hifluxbias}.  The impact of this bias on the summation of gas and stellar contributions to total baryonic mass are discussed in \S \ref{sec:relativemasses}.  Our treatment of the bias is validated through tests with a mock catalog in \S \ref{sec:mock-test}.  With the general properties of the BTFR now established, we carry out a Bayesian analysis to refine parameters, first at optical bands in \S \ref{sec:optimizeOP}, and then the coupling to the infrared in \S \ref{sec:optimizeIR}.

The core product of \S \ref{sec:btfr-formalism} is the formulation of the BTFR across three optical bands and one infrared band with an arbitrary absolute scale set by assuming Hubble expansion with $H_0=75$~\kmsMpc.  In \S \ref{sec:zp}, the scale is shifted (slightly) to match constraints imposed by BTFR values for samples of galaxies with Cepheid and Tip of the Red Giant Branch distance measurements.  Our method for determining distances to individual galaxies is discussed in \S \ref{sec:distMeasure}, with uncertainties the topic of \S \ref{sec:uncertaintyMeasure}.  The message in this latter section is that the sum of error estimates on observed and derivative parameters does not fully explain the BTFR scatter; ie, an unexplained or intrinsic component contributing to scatter is significant.

Given distance estimates for the $\sim 10,000$ galaxies in our sample, in \S \ref{sec:H0} we derive a best value for $H_0$ consistent with our data.  Various tests are carried out in \S \ref{sec:tests} that the sample must reasonably pass if there is to be confidence in the relative (not absolute) distances that are derived.  A tabulation of our results is presented in \S \ref{sec:table}. 
The BTFR distances derived here are a component of the ensemble of distances that will be assembled in an upcoming {\it Cosmicflows-4} publication. The absolute scale of the BTFR distances and associated determination of $H_0$ given here should be considered preliminary. 

\section{The BTFR Formalism}
\label{sec:btfr-formalism}

The baryonic version  of the TFR is described by the equation
\begin{equation}
    {\rm log} M_b = Slope ({\rm log}W^i_{mx}-2.5) + ZP,
\label{eq:BTFR}
\end{equation}
where baryonic mass $M_b$ replaces luminosity, whether optical or infrared. Here, $W^i_{mx}$ is the \hi line width corrected for inclination, $i$, following $W^i_{mx}=W_{mx}/{\rm sin}(i)$.
The baryonic mass is defined as the combination of gas and stellar mass through the equation
\begin{equation}
    M_b = K_g M_{HI} + \Upsilon^*_{\lambda} L_{\lambda}
    \label{eq:Mb}
\end{equation}
where $M_{HI}$ is the mass in neutral Hydrogen, 
$K_g$ is a multiplier that accounts for interstellar gas in forms other than neutral Hydrogen,
$L_{\lambda}$ is the absolute luminosity in the passband $\lambda$, and $\Upsilon^*_{\lambda}$ is the stellar mass-to-light ratio appropriate to the passband $\lambda$.

A substantial complexity arises because our sample ({\it any} large BTFR sample) is neutral Hydrogen flux limited, creating a bias that must be addressed.  We take an iterative approach to the problem.  Initially, we formulate the BTFR disregarding the bias and with simplistic assumptions.  We then demonstrate the bias and a procedure to build a pseudo-BTFR that annuls the bias for the purpose of measuring distances.  The procedure is tested against a mock catalog. The full formulation involves the parameters governing the translations from luminosities to stellar masses and \hi fluxes to gas masses, the \hi flux bias adjustment, and the slopes and zero points in the separate optical and infrared BTFR equations.  Solutions for these parameters are derived through a Bayesian Monte Carlo analysis. 

\begin{figure}
\centering
\includegraphics[width=0.95\linewidth]{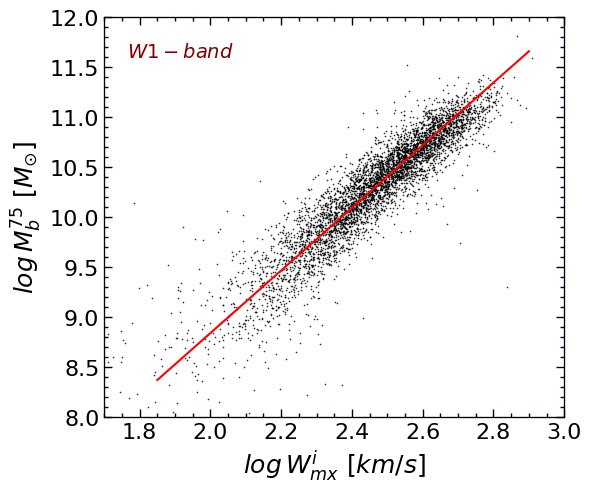}
\caption{Correlation between baryonic mass and \hi profile linewidth with WISE $W1$ photometry assuming $H_0=75$~\kmsMpc. 
The red line is the least-square fit with errors taken in the linewidth.  
}
\label{fig:WBTFR}
\end{figure}

\begin{figure*}
\centering
\includegraphics[width=1\linewidth]{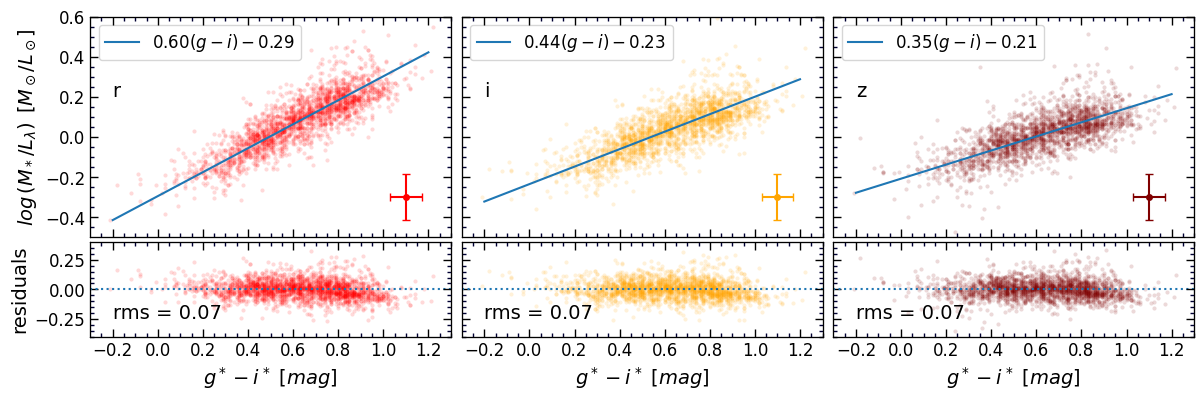}
\caption{
Stellar mass to light ratio versus color. Stellar mass, $M_*$, is determined using $W1$-band photometry assuming $\Upsilon^*_{W1} = 0.5~\Msun/\Lsun$, while the light is measured at $r$, $i$ and $z$ bands. The blue line is the best fit that is determined by minimizing the orthogonal distances of data points from the line. Here, we used Python package {\tt LtsFit} developed by \citet{2013MNRAS.432.1709C} that incorporates uncertainties on all variables, removes large outliers and accounts for the unknown intrinsic scatter.
The typical error bar of the scattered points is given at the bottom right corner of each panel in the top row.
\label{fig:linkopir}}
\end{figure*}


\subsection{Gas Mass}
\label{sec:gas}

The mass in neutral Hydrogen gas is \citep{1984AJ.....89..758H}
\begin{equation}
    M_{HI} = 2.36 \times 10^5 d^2 F_{21}
    \label{eq:MHI}
\end{equation}
where $F_{21}$ is the integrated flux in the 21cm \hi signal in Jy \kms\ and $d$ is the distance of the target in Mpc.  
Taking the factor $K_g=1.33$ in Equation~\ref{eq:Mb} accounts for the contribution from interstellar Helium that must accompany Hydrogen.  In the literature \citep{2016ApJ...832...11B} the value $K_g=1.4$ has been used to account for molecular, ionized, and metal enriched constituents of interstellar gas.  We have evaluated results based on these alternative gas multipliers and not found a significant effect.  We assume $K_g=1.33$. 

Values for the integrated \hi flux $F_{21}$ are drawn from two sources.  One is the {\it All Digital \hi} file posted in the {\it Extragalactic Distance Database}\footnote{\url{https://edd.ifa.hawaii.edu}} \citep{2009AJ....138.1938C, 2021A&A...646A.113D}.  This compilation for more than 20,000 galaxies draws on observations from the Green Bank 100-meter, 300-foot, and 140-foot telescopes, the Arecibo Telescope, the Parkes Telescope, the Nancay Telescope, and the Effelsberg Telescope.  All observations summarized in this file, whether originating from our own observations or from others, have been analyzed in a coherent fashion by drawing raw material from the various archives.  The other source is the {\it ALFALFA 100\%} compilation \citep{2018ApJ...861...49H}, available from the {\it Arecibo Legacy Fast ALFA Survey} website\footnote{\url{http://egg.astro.cornell.edu}} or at the {\it Extragalactic Distance Database}.\footnote{EDD fully includes material from the ALFALFA 40\% compilation \citep{2011AJ....142..170H} analyzed in a consistent way with data from other sources.}  An unweighted average is taken if fluxes are available from both sources.  In all cases, the observations are made with single beam radio telescopes.  In the vast majority of cases, the beams are substantially larger than the optical images of the targeted galaxies.  Flux could be lost with particularly extended gas-rich galaxies or with very nearby galaxies of large angular extent.

\begin{figure}
\centering
\includegraphics[width=0.95\linewidth]{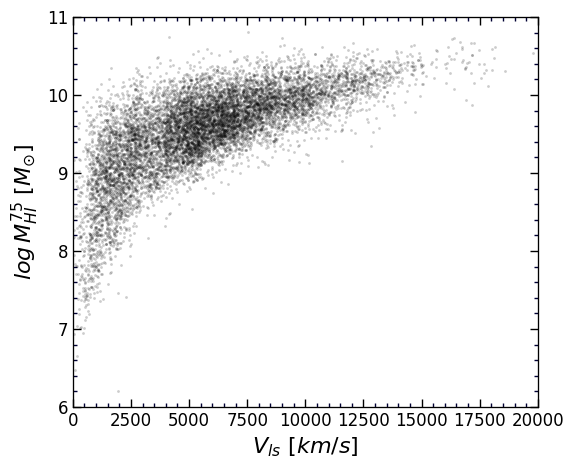}
\caption{Mass in \hi from \hi fluxes as a function of systemic velocity.
\label{fig:VHI}}
\end{figure}

\begin{figure}
\centering
\includegraphics[width=0.95\linewidth]{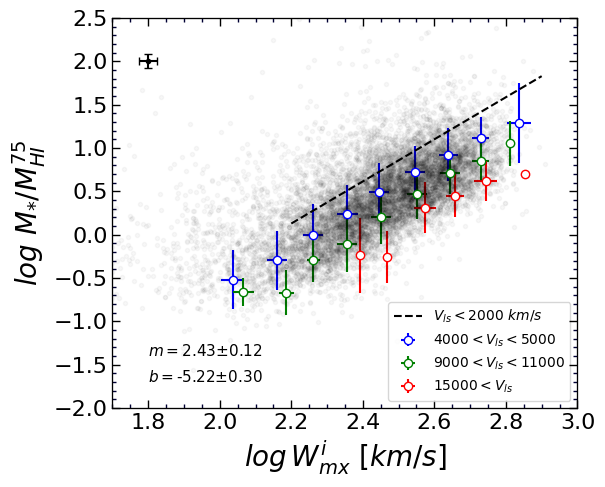} \\
\includegraphics[width=0.95\linewidth]{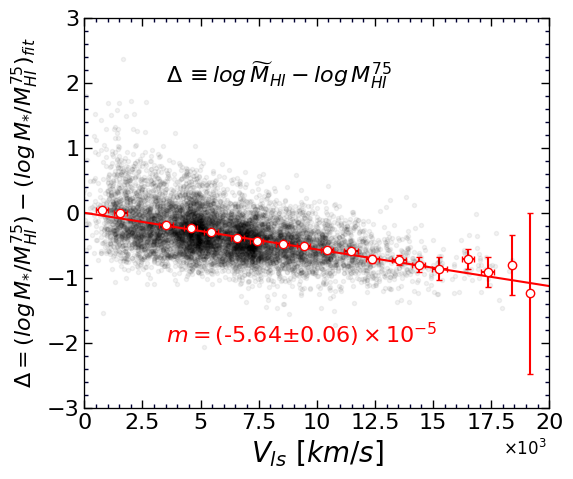} \\
\includegraphics[width=0.95\linewidth]{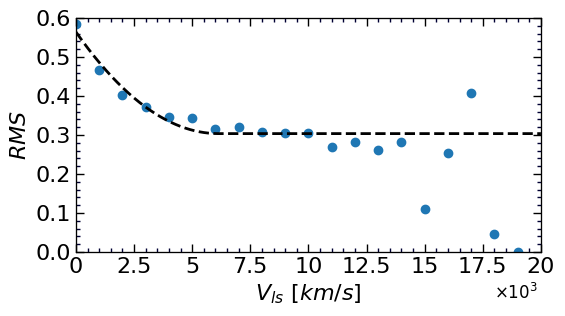} 
\caption{
{\bf Top: } The stellar to \hi gas ratio versus linewidth. Open symbols are the average of data points within bins of 0.1 in log~$W_{mx}^i$. The black dashed line is the least-square fit 
to galaxies with $V_{ls}< 2000$ \kms\ and $\rm{log}W^i_{mx}>2.2$ \kms. $m$ and $b$ are the slope and zero point of the black dashed line. The black error bars on the top left corner of this panel indicate the typical uncertainty of the black data points.
{\bf Middle:} The vertical deviation of black data points from the dashed line in the top panel versus radial velocity, $V_{ls}$. Open symbols represent the average of data points in velocity bins of 1000 \kms, and error bars show the $1\sigma$ scatter of the average values. The red solid line is the result of a least-square fit minimizing the residuals along the vertical axis with $m$ being the slope of the line.
{\bf Bottom:} The rms scatter of the black data points about the red linear correlation in the middle panel. Blue points represent the median scatter values within the velocity bins of $1000$ \kms. Black dashed line represents the least-square fitted curve, which has a quadratic form on the left side (small velocities) and remains constant after a specific velocity determined in the fitting process. 
\label{fig:delHI}}
\end{figure}
\subsection{Stellar Mass}
\label{sec:stellar_mass}

The relationship between stellar mass and luminosity has been extensively studied at infrared bands where flux is dominated by evolved stars on the red giant branch and optimal correlations are anticipated between light and mass.  Particular attention has been given to the L-band where observations have been carried out with the space telescopes Spitzer ([3.6] filter) and WISE (adjacent W1 filter).

Stellar population synthesis or star formation history models can inform the value and scatter of the parameter $\Upsilon^* = M^*/L$ at the closely similar [3.6] and W1 bands \citep{2012AJ....143..139E, 2014AJ....148...77M, 2014ApJ...782...90C, 2014ApJ...788..144M, 2015ApJS..219....5Q, 2018ApJ...865..154H, 2019MNRAS.483.1496S, 2019MNRAS.484.3267L}.  There is general agreement from models suggesting $\Upsilon^*_{W1} \simeq 0.5~\Msun/\Lsun$.  The choice of stellar initial mass function is a dominant source of systematic uncertainty.  As \citet{2015ApJ...802...18M} have pointed out, constraints on $\Upsilon^*_{W1}$ based on the minimization of scatter of the BTFR provides constraints on systematics in the stellar mass distribution models.

With an initial assumption $\Upsilon^*_{W1} = 0.5~\Msun/\Lsun$, with observed \hi and WISE $W1$ fluxes, and taking distances\footnote{Hubble flow distances are derived from velocities in the Local Sheet rest frame \citep{2008ApJ...676..184T}, assuming group mean velocities for galaxies identified to be in a group.  The Local Sheet frame is preferred over the cosmic microwave background frame nearby because of a coherence in galaxy flows that extends over at least 40~Mpc.  Velocities in alternate reference frames converge at large distances.} from the Hubble law assuming $H_0 = 75$~\kmsMpc, we generate Figure~\ref{fig:WBTFR}.  The value of $\Upsilon^*_{W1}$ is the single free parameter in this plot.
Scatter arises from variance in $\Upsilon^*_{W1}$ and a multitude of other sources (uncertainties in \hi fluxes, line widths, inclinations, photometry, and obscuration, distance deviations from the Hubble law, and intrinsic scatter).

At optical bands, where Population~I components compete with Population~II, $\Upsilon^*_{\lambda}$ have color terms \citep{2011MNRAS.418.1587T}.  The dependence with SDSS $g - i$ is shown in Figure~\ref{fig:linkopir} for the cases of the $r$, $i$ and $z$ bands.
On the ordinate, we plot stellar mass to light ratios (assuming Hubble flow distances based on $H_0 = 75$~\kmsMpc).
Here, where we require both optical and infrared information to be available, the stellar mass is based on photometry at the WISE $W1$ band, as discussed in the previous paragraph, while the absolute luminosity, $L_{\lambda}$, is derived from SDSS $\lambda=r$, $i$ or $z$ photometry.\footnote{All $SDSS$ optical and $WISE$ infrared photometry used in this study has been carried out from raw archival material and inclinations and obscuration issues have been given detailed attention as discussed exhaustively by \citet{2019ApJ...884...82K, 2020ApJ...896....3K, 2020ApJ...902..145K}}.  
This mixed photometry provides a linkage between the infrared and optical, assuring that $<\Upsilon^*_{\lambda} L_{\lambda}> \simeq <\Upsilon^*_{W1} L_{W1}>$.  The linkage is accomplished through the relations demonstrated by the blue lines in Figure~\ref{fig:linkopir}
\begin{equation}
    {\rm log}{\Upsilon^*_{\lambda} = {\rm log} (M_*/L_{\lambda})=\alpha_{\lambda}(g-i)+\beta_{\lambda} }.
\label{eq:colorterm}
\end{equation}
where $M_*$ is the stellar mass, and $\lambda$ is any of $r$, $i$ and $z$ bands. In this study, for the absolute magnitude of the Sun we adopt the values $4.64$, $4.53$, and $4.50$ mag in the AB system at SDSS $r$, $i$ and $z$ bands, respectively \citep{2018ApJS..236...47W}. Our derived color dependencies are in a good agreement with the predictions of stellar population models \citep{2003ApJS..149..289B, 2014AJ....148...77M}. 

A tentative distance to a galaxy is given by its deviation from the fit to the data in Figure~\ref{fig:WBTFR}.  
Scatter about the fitted line arises for many reasons beyond unknowns that could be called intrinsic. Horizontal scatter arises from uncertainties in line width measurements and inclination adjustments.  Vertical scatter arises from photometry errors (small), obscuration (small in the infrared), variations in $\Upsilon^*_{\lambda}$, and uncertainties in \hi flux.  Vertical scatter also arises from departures in distance from the expectations of uniform cosmic expansion.  These distance components have a unique signature.  Galaxies at distance $d$ that are greater than given by the Hubble law are intrinsically brighter than assumed, whence $M^d_b > M^{75}_b$ where these galaxies are given representation by $M^{75}_b$ in Figure~\ref{fig:WBTFR}.  Such galaxies lie lower in this figure than they would if baryonic mass on the ordinate was based on their correct distance.  The inverse is the case for galaxies nearer than distances given by the Hubble law; the baryonic masses of such galaxies will lie high with respect to what their placement would be if they were given correct distances.  

If the only reason for scatter in the BTFR were variations in distance at a given systemic velocity then the distance for a galaxy, $j$, with respect to the arbitrary $H_0=75$ scale would be 
\begin{equation}
  d_j/d_{75}=(M^{fid}_{b,j}/M^{75}_{b,j})^{1/2},
  \label{eq:d_est}
  \end{equation}
where $d_{75}$ is the distance given by the Hubble law, and $M^{fid}_{b,j}$ is the baryon mass on the fiducial relation at the observed linewidth of the galaxy.
Of course, only a small part of the scatter in the BTFR is attributable to distance deviations; indeed, is a component that diminishes roughly linearly with redshift.  Assuming the scatter that is not associated with distance variations is randomly distributed about the fiducial relation, then measures of distances are embedded in deviations from the fiducial relation as systematics but buried in the substantial random uncertainties of non-distance causes.
A more sophisticated recovery of distance estimates than offered by the simple application of Eq.~\ref{eq:d_est}  will be described in \S \ref{sec:distMeasure}.

\subsection{\hi Flux Bias} \label{sef:HI_Flux_Bias}
\label{sec:hifluxbias}

Although the BTFR seen in Figure~\ref{fig:WBTFR} looks promising, it contains a bias that is revealed by considering the run of the Hubble parameter 
\begin{equation}
H_j = f_j V_{ls,j} / d_j 
\label{eq:hubbleparameter}
\end{equation}
given systemic velocity, $V_{ls,j}$, for galaxy $j$ with distance $d_j$.  The function $f_j$ accounts for cosmic curvature\footnote{$f_j=1+1/2(1-q_0)z_j-1/6(1-q_0-3q_0^2+j_0)z_j^2$ where $z_j$ is the redshift of the galaxy, $j_o\simeq1$, and $q_0=0.5(\Omega_m-2\Omega_{\Lambda})=-0.595$ assuming $\Omega_m=0.27$ and $\Omega_{\Lambda}=0.73$. \citep{2004CQGra..21.2603V}} which is slight within the range of the current sample. The bias is in the sense that $H_j$ values drift lower (distances tending to overestimation) with increasing redshift.  The bias arises because the primary selection of targets comes from \hi surveys that are flux limited, particularly the dominant ALFALFA survey \citep{2018ApJ...861...49H}.  See Figure~\ref{fig:VHI} and  \citet{2021A&A...646A.113D} Figure 3.  Photometry is comfortably available for any galaxy that meets the \hi detection criterion.  Nearby, \hi is easily detected in any galaxy relevant to our interests.  However, at increasingly great redshifts galaxies with lesser \hi flux drop out of the sample.  The consequence is manifested in the top panel of Figure~\ref{fig:delHI} which plots the differential stellar and gas components vs. linewidth overlain with binned differences in increasing redshift intervals.  It is clear that the measured \hi component of mass becomes increasingly important at larger redshifts.  The observed ratio $M_{\star}/M_{HI}$ decreases with increasing systemic velocity.

The bias against candidates with lesser \hi flux is given quantitative expression in the middle panel of Figure~\ref{fig:delHI}. The difference, $\Delta$, gives the departure from the dashed straight line at a given ${\rm log}W^i_{mx}$ in the top panel.  Letting $\Delta \equiv {\rm log}\widetilde{M}_{HI}-{\rm log}{M}_{HI}$, the linear relation in the middle panel is expressed as
\begin{equation}
{\rm log}\widetilde{M}_{HI} = {\rm log}{M}_{HI}-5.64\pm0.06\times 10^{-5}~V_{ls}.
\label{MHIcorrection}
\end{equation}
What we will call the pseudo-gas mass parameter, $\widetilde{M}_{HI}$, is reduced from observed $M_{HI}$ values such that $<M_{\star}/\widetilde M_{HI}> \simeq constant$ with redshift. With this new parameter we can formulate a pseudo-baryonic mass, $\widetilde{M}_b$, that obeys a pseudo-BTFR 
\begin{equation}
    {\rm log} \widetilde{M}_b = Slope ({\rm log}W^i_{mx}-2.5) + ZP.
\label{eq:cBTFR}
\end{equation}
In \S \ref{sec:mock-test}, it will be demonstrated with mock data that unbiased distance estimates can be recovered based on Equation~\ref{eq:cBTFR}.

\begin{figure}
\centering
\includegraphics[width=0.95\linewidth]{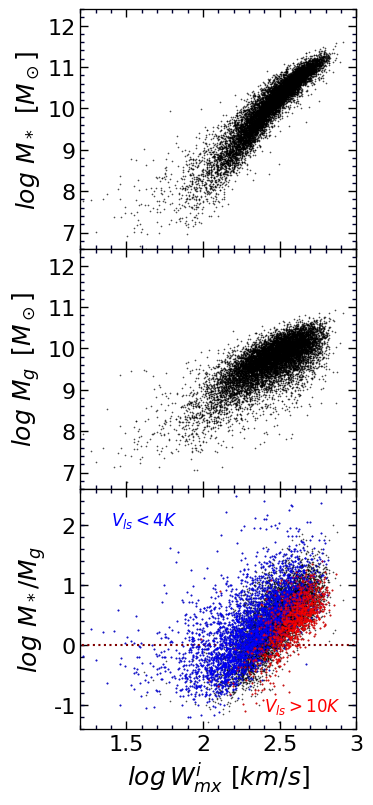}
\caption{Separate mass components. Top: Stellar mass vs. linewidth for entire sample. Middle: Gas mass vs. linewidth for entire sample. Bottom: Ratio of stellar mass to gas mass for galaxies with velocities less than 4,000~\kms\ (blue) and greater than 10,000~\kms\ (red). Stellar mass dominates in cases above the dotted line in the bottom panel; gas mass dominates in cases below the dotted line. 
\label{fig:3masses}}
\end{figure}

\begin{figure*}
\centering
\includegraphics[width=1.0\linewidth]{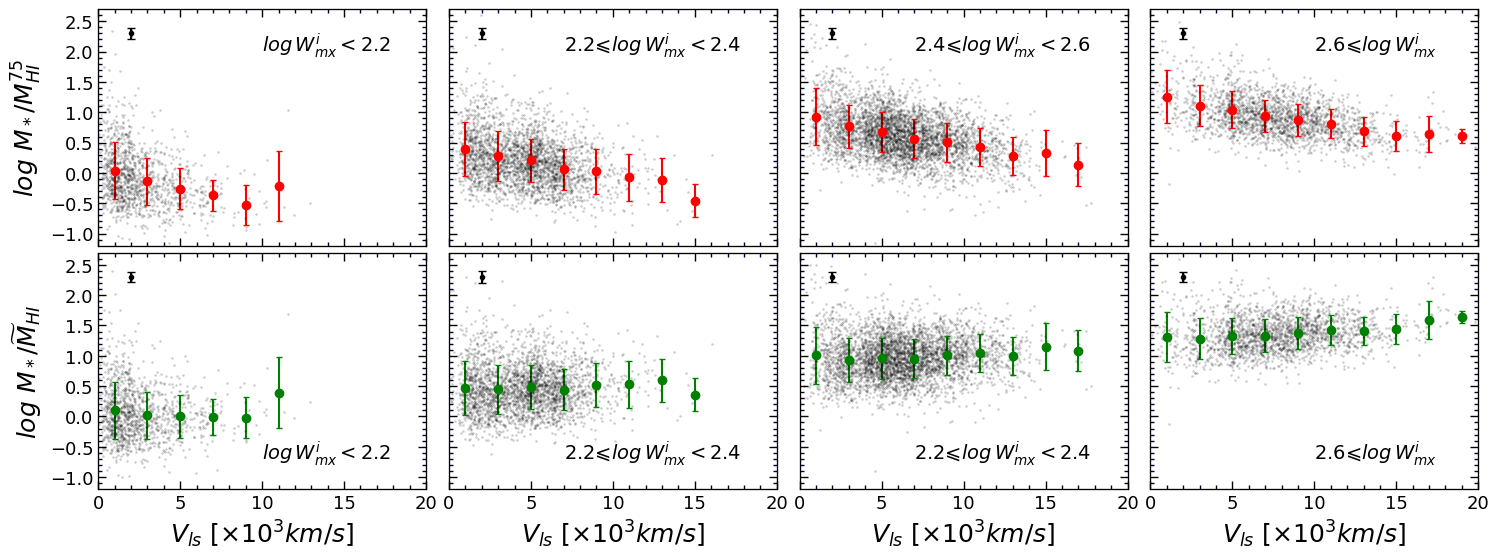}
\caption{
The stellar to \hi gas ratios versus systemic velocity, within different linewidth ranges. Raw and adjusted \hi gas mass are used to generate the top and bottom panels, respectively.
Each black point represents a galaxy. Filled symbols are the average of data points within the velocity bins of $2000$ \kms, with error bars indicating $1\sigma$ scatter within each bin. The black error bars on the top left corner show the typical error bar of the black dots.
\label{fig:Ms_Mhi_vs_Velocity_WimxBins}
}
\end{figure*}

\subsection{Relative Stellar and Gas Mass contributions}
\label{sec:relativemasses}

The separate mass contributions from stars and gas are shown as a function of \hi profile linewidths in the top two panels of Figure~\ref{fig:3masses}.  The bottom panel shows the ratio of the two for selections of the sample.  The stellar mass has by far the tighter correlation with rotation.  Gas masses peak abruptly at $M_g \simeq 4\times10^{11}~\Msun$ while stellar masses reach $M_* \simeq 2\times10^{12}~\Msun$.  Gas masses are relatively unimportant for the largest galaxies.

In the bottom panel, a variant of the top panel of Figure~\ref{fig:delHI}, stellar to total gas mass ratios are shown for galaxies at the velocity extremes of our sample: those with velocities less than 4,000~\kms\ are identified in blue while those with velocities greater than 10,000~\kms\ are identified in red.  The displacement toward lower $M_*/M_g$ values at higher velocities, the consequence of the \hi flux bias, is evident.  However another point can be made.  The horizontal dotted line separates between an upper zone of stellar mass dominance and a lower zone of gas mass dominance.  Our sample includes all manor of galaxies within the low velocity (distance) cutoff of 4,000~\kms. Galaxies with large systemic velocities (distances) such as those represented in red in the figure are massive galaxies with the stellar component dominant.  This situation somewhat mitigates the \hi flux bias problem.  The \hi detection bias is aggravated at large distances but most sample targets at large distances have only minor gas contributions.  Nearby, the gas contributions can be substantial but well formulated because the bias is small.

The redshift dependency of the stellar to \hi gas ratios is also evident in the top row panels of Figure \ref{fig:Ms_Mhi_vs_Velocity_WimxBins}. 
Several trends are evident. For one, there is an upward shift in stellar to \hi mass with increasing linewidths. Second, the preponderance of samples shifts to higher velocities with increasing linewidth intervals.  But third and a direct consequence of the \hi flux bias, there are downward trends in stellar to \hi mass ratios with systemic velocity across all the upper panels.
The bottom row panels of Figure \ref{fig:Ms_Mhi_vs_Velocity_WimxBins} are generated in a similar fashion using the \hi pseudo-mass, $\widetilde{M}_{HI}$. The consequence of the adjustments is a flattening of the slope of the stellar to \hi gas mass ratios with redshift. Utilizing $\widetilde{M}_{HI}$ minimizes the redshift dependent scatter about the pseudo-BTFR and results in distances free of the \hi flux bias.

\subsection{A Mock Catalog Test}
\label{sec:mock-test}

The adjustments made to compensate for the bias caused by the \hi flux cutoff can be evaluated through the study of a mock catalog of objects with known distances and otherwise similar observed properties as our real sample.  We construct a mock catalog with the following recipe.
(1) Create a sample that populates the fiducial BTFR assuming distances based on $H_0=75$~\kmsMpc\ and assuming a Schechter function that is appropriate for \hi selection \citep{2020ApJ...902..145K}.  The construction assumes the gas to stellar mass fraction given by \citet{2012MNRAS.425.2741H} with the scatter prescribed in that paper.
(2) Create Gaussian scatter in both log linewidth and log baryon mass. 
(3) Translate this sample to a multitude of distance shells so now the operational parameters are apparent magnitudes and observed \hi fluxes.  Populate the shells such that after imposing an \hi flux limit there is rough agreement with the histogram with redshift of the observed sample.
(4) Only retain candidates that exceed an \hi flux limit.
(5) Determine the distances that would be measure for the candidates that are retained and compare those distances with the known input distances.

\begin{figure}
\centering
\includegraphics[width=0.95\linewidth]{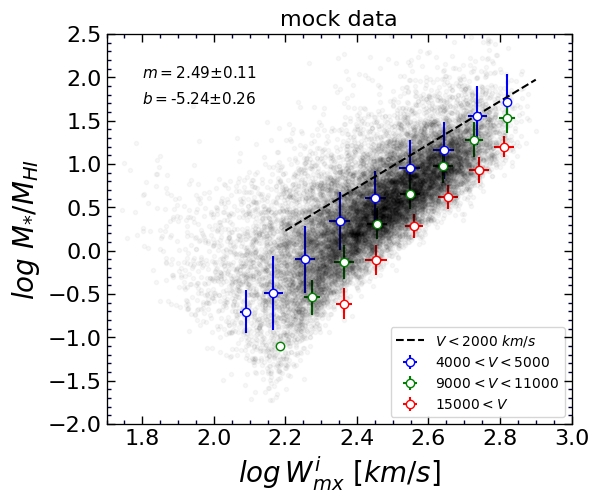} \\
\includegraphics[width=0.95\linewidth]{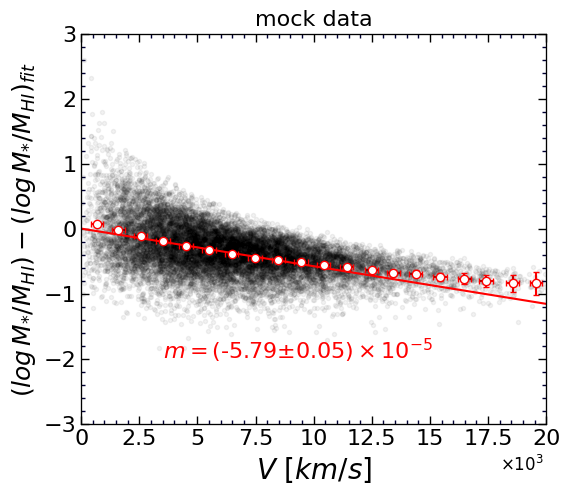} 
\caption{
{\bf Top: } The stellar to \hi gas ratio versus linewidth similar to the presentation in the top panel of Fig.~\ref{fig:delHI} but for the mock data.  The symbols and dashed lines have the same representations in the two figures.
{\bf Bottom:} Vertical offset of mock data from the dashed line of the left panel as a function of systemic velocity. The pattern seen in the middle panel of Fig.~\ref{fig:delHI} is recovered.
\label{fig:mockdelHI}}
\end{figure}

The analysis of the mock material follows what was done with the real data.  The top panel of Figure~\ref{fig:mockdelHI} shows the stellar to gas mass fractions as a function of linewidths with superimposed symbols giving binned averages within systemic velocity intervals.  In the bottom panel of this figure, individual points record the vertical position of a candidate with respect to the dashed line in the top panel as a function of systemic velocity.  Mean values in velocity intervals are represented by open symbols and a linear fit is given to the distribution.  The slope of this line of $-5.79\pm0.05 \times 10^{-5}$ is very close to that   of Eq.~\ref{MHIcorrection} describing the real data. 

\begin{figure}
\centering
\includegraphics[width=0.95\linewidth]{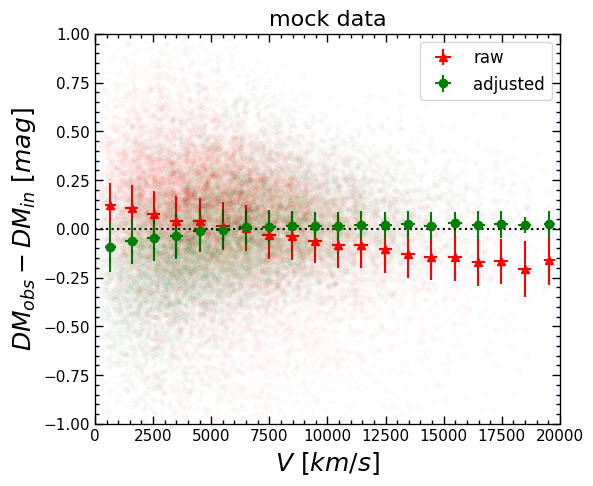} 
\caption{
Deviations of recovered (observed) distance moduli from input moduli for the mock data as a function of systemic velocity.  The directly measured (raw) moduli are coded in red and the adjusted values are in green.
\label{fig:mockdelD}}
\end{figure}

\begin{figure}
\centering
\includegraphics[width=0.95\linewidth]{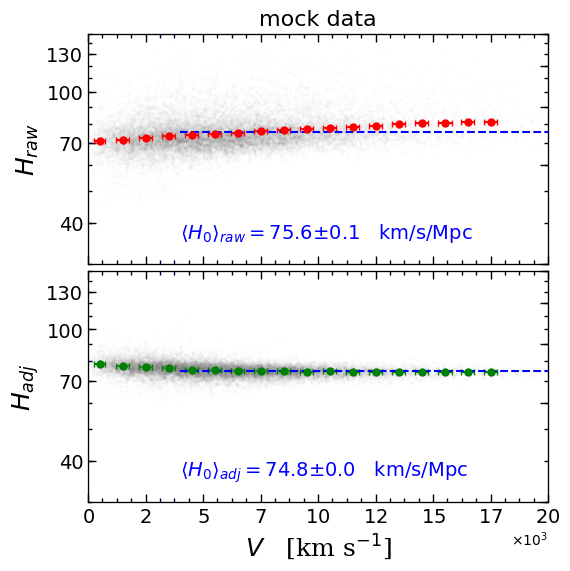} 
\caption{
Hubble parameter vs systemic velocity for the mock data for directly measured values in the top panel and the adjusted values in the bottom panel.
\label{fig:mockHV}}
\end{figure}

The fit in the bottom panel of Figure~\ref{fig:mockdelHI} provides the information needed for adjustments to mock distance moduli.  Figure~\ref{fig:mockdelD} illustrates the results with respect to input values scattered around $H_0=75$~\kmsMpc\ for the raw observed moduli in red and the adjusted moduli in green.  The raw observed moduli drift to higher distances at larger velocities as expected from the bias.  The adjustment returns distances that are close to the mean values that would be recovered if there were no \hi flux cutoff.  These same results are seen in Figure~\ref{fig:mockHV} where the ordinate is now the Hubble parameter $V_j/d_j$, with means in velocity bins and horizontal fits represented by the dashed lines for values at $V>4000$~\kms.  In the lower panel adjusted case, the input value of $H_0$ is recovered and beyond 4,000~\kms\ the averaged values are stable about constant $H_0$.

\begin{figure*}
\centering
\includegraphics[width=0.90\linewidth]{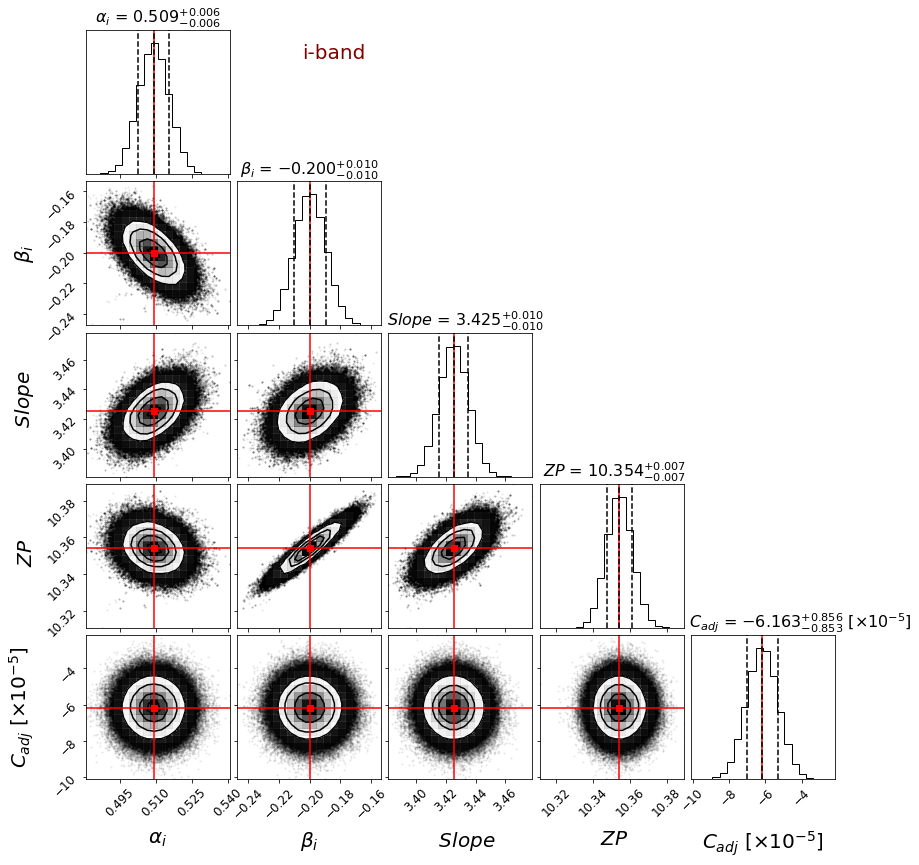} 
\caption{
The posterior distribution of the optimized parameters of the BTFR model as described in Equations \ref{eq:cBTFR} and \ref{eq:colorterm}, with $i$-band photometry as the primary source of the derived stellar mass.
Contours represent $\sigma/2$, $\sigma$, $3\sigma/2$ and $2\sigma$ levels of the 2-dimensional distributions and they enclose 12\%, 39\%, 68\% and 86\% of the distributed points, respectively.
\label{fig:mcmc_i_gi_rz}}
\end{figure*}

\begin{figure*}
\centering
\includegraphics[width=0.32\linewidth]{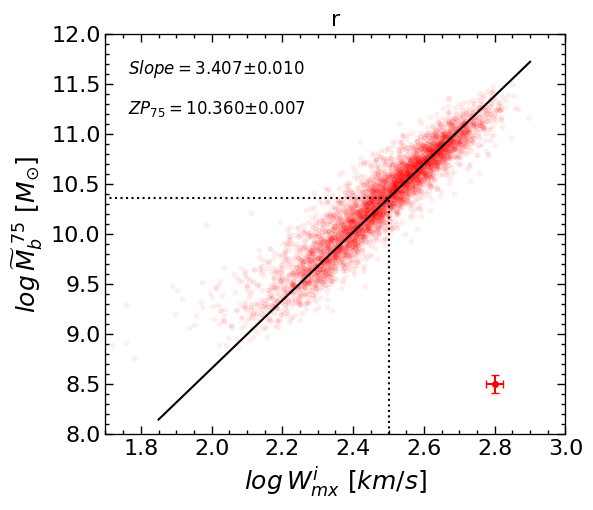} 
\includegraphics[width=0.32\linewidth]{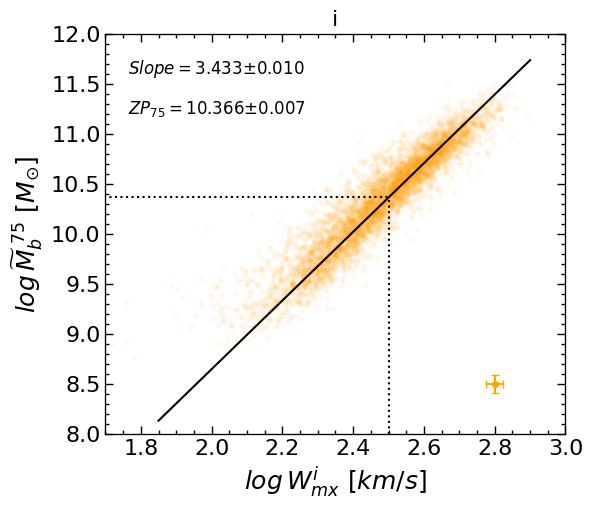} 
\includegraphics[width=0.32\linewidth]{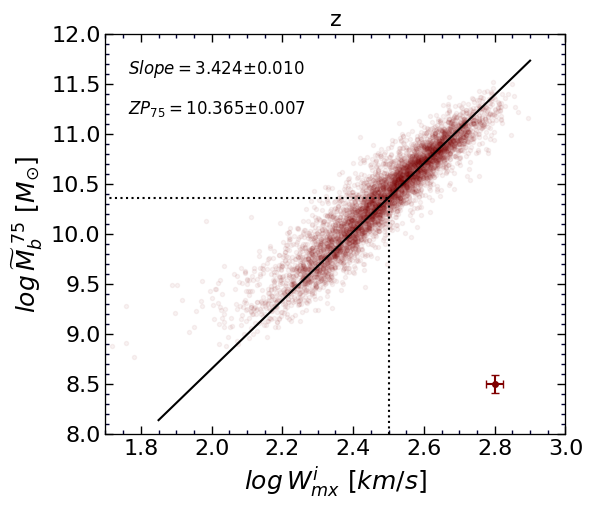} 
\includegraphics[width=0.32\linewidth]{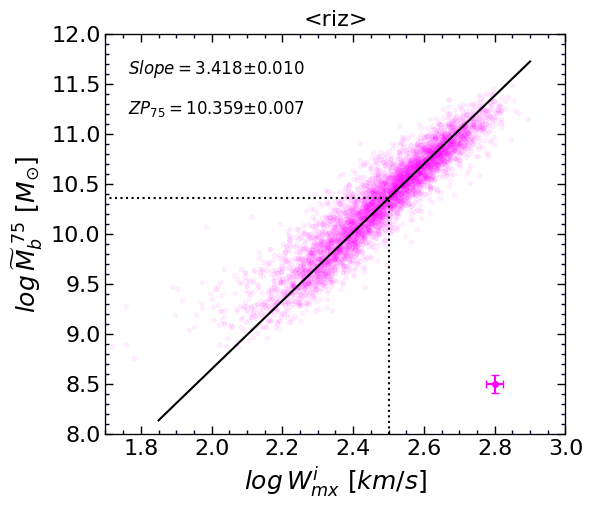} 
\includegraphics[width=0.32\linewidth]{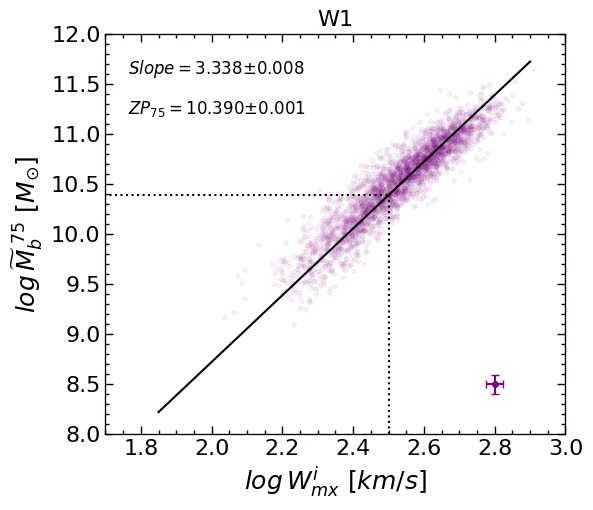} 
\caption{
BTFR after compensation for the \hi selection bias using $\widetilde{M}_b=1.33\widetilde{M}_{HI}+M_*$.  Black lines illustrate fits with errors taken in linewidths. 
The slopes and zero points referenced to log $W^i_{mx} = 2.5$ are given in the legends. The BTFR models are based on stellar mass to light ratios given by ${\rm log} (M_*/L_{\lambda})=\alpha_\lambda(g-i)+\beta_\lambda$ (see Table \ref{tab:mcmc-params}).
{\bf Top:} BTFR with SDSS $r$ photometry (left), $i$ photometry (middle) and $z$ photometry (right) after compensation for the \hi selection bias.
Bottom: Similar to the top row but for BTFR with SDSS averaged $r$, $i$ and $z$ photometry (left) and WISE infrared $W1$ photometry (right).  Error bars in the bottom right corner of each panel show the typical uncertainty on the scattered points.
\label{fig:BTFR-nonCalibrated}}  
\end{figure*}


\subsection{Rigorous Parameter Constraints: Optical Bands}
\label{sec:optimizeOP}

We carried out SDSS photometry at $u,g,r,i,z$ bands \citep{2020ApJ...896....3K, 2020ApJ...902..145K}. The $u$ material will not be considered further because measurement uncertainties are larger in that band. The TFR at $g$ has increased scatter because of stochastic contributions from young populations.  The utility of $g$ photometry is its service in identifying such contributions, as seen in Figure~\ref{fig:linkopir}. On the other hand, the TFR at $r$, $i$ and $z$ bands have comparable scatter, although color terms are slightly different.

Our aim in this section is to explore the space of the model free parameters to find the best set of parameters that minimize the scatter about the linear relation between baryonic mass and \hi linewidth in Equation~\ref{eq:cBTFR}
where $\widetilde{M}_b$ is the pseudo-baryonic mass and is derived using the adjusted \hi mass, ${\rm log}\widetilde{M}_{HI} = {\rm log}{M}_{HI}-C_{adj}V_{ls}$. Here, $C_{adj}$ is the adjustment factor and one of the model parameters. In this section, we relax the assumption we made in Section \ref{sec:btfr-formalism} about the non-variability of $\Upsilon^*_{W1}$ across all of our sample. 
We attempt to initially determine the functional form of the $\Upsilon^*_{\lambda}$ at optical wavebands and then find an equivalent version at infrared bands. 
Accordingly, we define $\Upsilon^*_{\lambda}$ as a linear function of the optical colors $g-i$ given by Equation \ref{eq:colorterm}.

In our optimization process, the objective is to construct the posterior distribution of the model parameters, $\Pc(\Theta|\mathcal{D})$, where $\Theta$ is the vector that holds all model parameters (i.e. $Slope$ and $ZP$ of the BTFR, $\alpha_\lambda$ and $\beta_\lambda$, and the bias factor $C_{adj}$. 
The observed data, $\mathcal{D}$, consist of linewidths and galaxy luminosities at different wavebands. 

The posterior distribution can be determined following the conditional probability law, $\Pc(\Theta|\mathcal{D}) \propto \Pc(\mathcal{D}|\Theta)\Pc(\Theta)$. We set $\Pc(\Theta)=1$ given a lack of any prior knowledge on the distribution of the model parameters.
This assumption reduces our problem to finding a set of optimum parameters that maximize the likelihood function. 

Having adopted the likelihood function we take advantage of the Markov Chain Monte Carlo (MCMC) method to explore the parameter space and sample the posterior distribution. This process allows us to incorporate all uncertainties of the observables upon which our model is constructed. Data points are mutually independent since they represent different galaxies. Moreover, it is a reasonable assumption that all uncertainties on the measured observables are almost Gaussian implying the following likelihood function
\begin{equation}
\label{Eq:likelihood}
\Lc= \prod_{n=1}^{N} \frac{1}{\sqrt{2 \pi \sigma_n^2}} ~\exp\Big(\frac{\Dc_n-\Mc_n(\Theta)}{\sigma_n}  \Big)^2 ~,
\end{equation}
where $n$ is the galaxy index and $N$ is the total number of galaxies in our analysis. 
To minimize the effect of the Malmquist bias, the data$-$model residuals ($r_n = \Dc_n-\mathcal{M}_n(\Theta)$) are calculated along the linewidth parameter \citep{1995PhR...261..271S,2000ApJ...533..744T}. 
Here, $\sigma_n$ is the uncertainty of the residual parameter, $r_n$, which penalizes a model based on the uncertainty of the real measurements and predictions, and is calculated using $\sigma_n^2=\sigma_{D_n}^2+\sigma_{\Mc_n}^2$. For each galaxy $\Dc_n={\rm log}W^i_{mx,n}$ and $\sigma_{D_n}$ is the uncertainty on the measured linewidths. For any model with given parameters, $\Theta$, the model uncertainty $\sigma_{\Mc}$ is calculated through a propagation of uncertainties of the model observables, i.e. \hi flux, magnitudes, and color indices. 

In this study, we sample the posterior distribution using Python package {\it emcee} \citep{2013PASP..125..306F},  {\it emcee} leverages fast linear algebra routines to accelerate the process and takes the logarithm of the posterior likelihood 
\begin{equation}
\label{Eq:likelihood1}
\log \Lc({\bf r})= -\frac{1}{2}{\bf r}^T\Sigma^{-1}{\bf r}-\frac{1}{2}\log |\Sigma|-\frac{N}{2}\log (2 \pi) ,
\end{equation}
where ${\bf r}$ is the $N\times1$ data$-$model residual as define earlier, and $\Sigma$ is the $N\times N$ diagonal covariance matrix where $\Sigma_{n,n}=\sigma^2_n$.

We adopt this likelihood for each of $r$, $i$ and $z$ bands and generate 64 chains each with the length of 10,000 samples. Each chain initially starts from a random location in the parameter space. This allows to expand the exploration area and obtain more robust results. 
Mostly after the first 500 burn in steps, each chain converges and follows the Markov statistics.
However, to be more conservative, we remove the first 1,000 steps of each MCMC sample and combine all 64 walkers to generate the posterior distribution of model parameters. 

The corner plot for the posterior distribution of model parameters based on the $i$ band photometry is illustrated in Figure \ref{fig:mcmc_i_gi_rz}.  In this diagram, the top-most panel of each column displays the one dimensional distribution of the corresponding sampled parameter, with the overplotted red solid line representing the median values, and the lower/upper bounds corresponding to 16/84 percentiles (black dashed lines). Horizontal and vertical red lines are drawn at the location of the median values that we adopt as the optimum values of our model parameters. The shape of the posterior distribution looks almost the same if either $r$ or $z$ band photometry is used to build the BTFR. 

\begin{table*}[t]
\centering
\caption{Optimized parameters of the BTFR and $\Upsilon^*_{\lambda}$ model.
\label{tab:mcmc-params}}

\begin{tabular}{c ccccc  c}
\hline \hline
waveband ($\lambda$) & $\alpha_\lambda$ & $\beta_\lambda$ & $Slope$ & $ZP_{75}^\dag$ & $ZP_{calib}^\ddag$ & $C_{adj}$ $[\times10^{-5}]$ \\
\hline 
$r$ & $0.650\pm0.006$ & $-0.243\pm0.010$ & $3.407\pm0.010$ & $10.360\pm0.007$ & $10.407\pm0.032$ & $-6.17\pm0.21$ \\
$i$ & $0.509\pm0.006$ & $-0.200\pm0.010$ & $3.433\pm0.010$ & $10.366\pm0.007$ & $10.411\pm0.033$ & $-6.16\pm0.86$ \\
$z$ & $0.461\pm0.007$ & $-0.190\pm0.010$ & $3.424\pm0.010$ & $10.365\pm0.007$ & $10.411\pm0.033$ & $-6.01\pm0.86$ \\
$<riz>$ & $0.536\pm0.010$ & $-0.206\pm0.010$ & $3.418\pm0.010$ & $10.359\pm0.007$ & $10.405\pm0.033$ & $-6.12\pm0.85$ \\
$W1$ & $-$ & $-$ & $3.338\pm0.008$ & $10.390\pm0.001$ & $10.408\pm0.026$ & $-6.20\pm0.82$ \\
\hline
\multicolumn{6}{l}{$^\dag$ Zero point based on distances assuming $H_0=75$~\kmsMpc.} \\
\multicolumn{6}{l}{$^\ddag$ Zero point after calibration based on distances from TRGB and Cepheid methodologies.}
\end{tabular}
\end{table*}

Table \ref{tab:mcmc-params} lists the optimized parameters of our model with the use of different optical bands and color combinations to calculate the stellar to mass ratios (Equation \ref{eq:colorterm}). The most probable $Slope$ and $ZP$ that we find through the MCMC simulations are in good agreement with each other at different wavebands. We tested that adding an extra linear dependency on the $r-z$ color term does not significantly alter the BTFR parameters.

The top row panels of Figure~\ref{fig:BTFR-nonCalibrated} illustrate the BTFR after adjustments for the \hi flux selection bias, for $r$, $i$ and $z$ bands, where for each waveband the model parameters are given in the corresponding row for the band in Table~\ref{tab:mcmc-params}. 


Our tools for measuring galaxy distances can be slightly sharpened by using a combination of $r,i,z$ material.
This unification can be accomplished in two ways. One is to extract the stellar mass of galaxies from the best fits of $\Upsilon^*_\lambda$ at different wavebands (Table \ref{tab:mcmc-params}). The other way is define an average linear relation 
\begin{equation}
    {\rm log}\Upsilon^*_{riz} = {\rm log} (M_*/L_{riz})=\alpha_{riz}(g-i)+\beta_{riz}
\label{eq:colortermRIZ}
\end{equation}
where $L_{riz}$ is the average of the absolute luminosities of a galaxy at $r$, $i$ and $z$ bands. We fit the parameters of this relation together with the BTFR coefficients through the same MCMC procedure presented in Section \ref{sec:optimizeIR}. Table~\ref{tab:mcmc-params} lists the output of our analysis and the bottom left panel of Figure \ref{fig:BTFR-nonCalibrated} illustrates the fitted straight line of the corresponding BTFR model.

\subsection{Extension to the Infrared Band}
\label{sec:optimizeIR}

The same methodology can be used to find model parameters at the infrared WISE $W1$ band. However, optical color indices are not available for galaxies in our sample that only have infrared photometry coverage. Therefore, we need a formulation that only uses infrared information to establish empirical functions involving $\Upsilon^*_{W1}$.  At the same time, though, we need to maintain a coherence between the optical and infrared measurements.  Consequently, in deriving the functional form of the infrared parameter $\Upsilon^*_{W1}$, we restrict to galaxies with both optical and infrared data coverage. We derive their stellar masses, $M^*_{riz}$, following Equation \ref{eq:colorterm} and the optimized parameters listed in Table \ref{tab:mcmc-params} using combined $<riz>$ optical measurements. 
The optical$-$infrared coupling is achieved by requiring
\begin{equation}
<\Upsilon^*_{riz} L_{riz}> \simeq <\Upsilon^*_{W1} L_{W1}>
\label{eq:op_ir_coupled}
\end{equation}
which results in $M^*_{riz}/L_{W1} \simeq \Upsilon^*_{W1}$.

\begin{figure}
\centering
\includegraphics[width=0.99\linewidth]{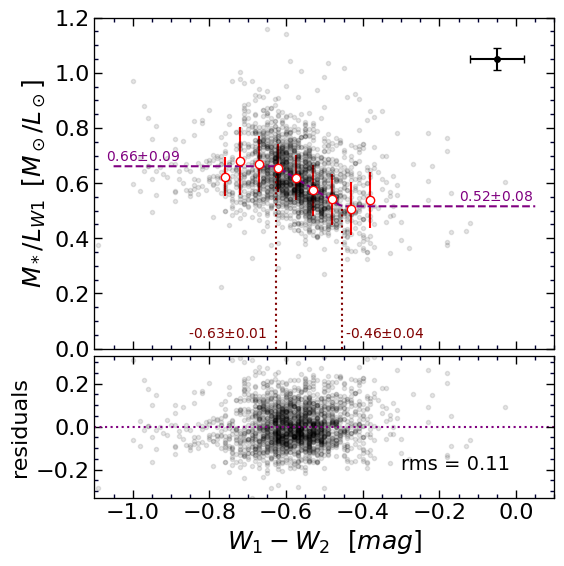} 
\caption{
{\bf Top:} $\Upsilon^*_{W1}$ versus $W1-W2$ color for galaxies with both optical and infrared data. Each black point represents a galaxy. Open red points show the average of data within bins of $0.05$ mag, with error bars showing the $1\sigma$ scatter of data points.
Purple dashed line represents the best model that is constructed using two constant color independent regions and a linear correlation in the middle. Numerical labels and the vertical dotted lines illustrate the boundaries of the slant line.  Black error bars on the top right corner show the typical uncertainty of the black points.
{\bf Bottom:} Residuals after subtracting the multi-linear model, displayed by the dashed line in the upper panel. The root mean square of residuals, rms, translates to $\sim19\%$ uncertainty in the values of $M_*/L_{W1}$.
}
\label{fig:ms_w1_w2}
\end{figure}

 Figure \ref{fig:ms_w1_w2} plots the optical stellar mass to $W1$-band luminosity versus the $W1-W2$ color. 
Compared to $g-i$ color with a dynamical range of $\sim 1.5$ mag, $W1-W2$ of most galaxies falls in a very short range $\sim 0.2$ mag. Still, a reasonable correlation pattern is identified when we bin data points. $M^*_{riz}/L_{W1}$ decreases toward redder $W1-W2$ colors at intermediate $W1-W$ values.  Statistics at blue and red extremes are poor where dependencies are taken to be approximately constant. The purple dashed line in Figure \ref{fig:ms_w1_w2} shows the best fit given by the maximum likelihood method. $M_*/L_{W1}$ takes the constant values of $0.66\pm0.09$ on the blue side of $W1-W2=-0.63\pm0.01$ mag. As the $W1-W2$ color gets redder, $M_*/L_{W1}$ linearly decreases with color until reaching the value of $0.51\pm0.08$ at $W1-W2 =-0.46\pm0.04$ mag. Redward, the value is taken to be constant at $0.52\pm0.08$. 

The scatter about this formulation is shown in the lower panel of Figure~\ref{fig:ms_w1_w2}.  The rms scatter in $M^*/L_{W1}$ of $0.11$ about a median value of $0.59$ corresponds to $19\%$ uncertainty on the stellar mass to light ratios.

\begin{figure}
\centering
\includegraphics[width=0.99\linewidth]{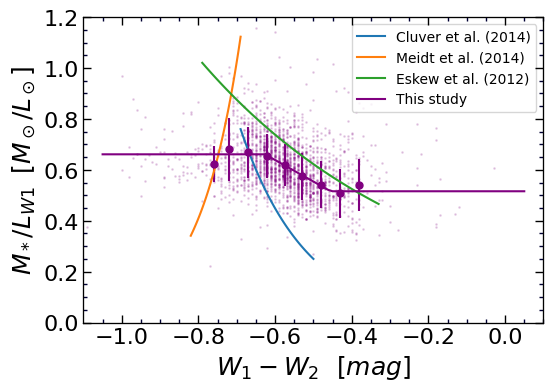} 
\caption{
Similar to Figure \ref{fig:ms_w1_w2}. The solid purple line traces $\Upsilon^*_{W1}$ as empirically derived in this study. The blue, orange and green curves are the stellar transformation functions of $\Upsilon^*_{W1}$, developed by \citet{2014ApJ...782...90C}, \citet{2014ApJ...788..144M} and \citet{2012AJ....143..139E}, respectively, over the range of their applicability. 
All magnitudes and colors are in the AB system. 
}
\label{fig:ms_w1_w2_literature}
\end{figure}

Figure \ref{fig:ms_w1_w2_literature} shows our model of $M_*/L_{W1}$ together with three theoretical models that are constructed by adopting various initial mass functions and stellar evolution scenarios \citep{2018ApJ...865..154H}. These models describe different galactic environments and are only valid for restricted ranges of $W1-W2$. 
As seen, our fitted function for the $M_*/L_{W1}$ color dependency (purple line) is in general consistent with those advocated by \citet{2014ApJ...782...90C}, and \citet{2012AJ....143..139E}. Most of our data points are located on the right side of the applicability zone of the \citet{2014ApJ...788..144M} curve. Our constraints on $M_*/L_{W1}$ at both $W1-W2$ color extremes are weak.

\begin{figure*}
\centering
\includegraphics[width=0.65\linewidth]{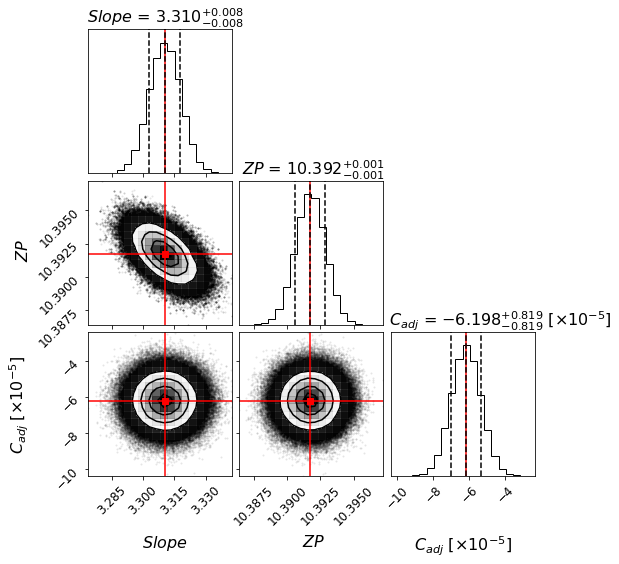} 
\caption{
The posterior distribution of the $Slope$ and $ZP$ of the BTFR based on the $W1$-band photometry, with $M_*/L_{W1}$ given the fitted multi-linear form presented in Figure \ref{fig:ms_w1_w2}. Other details are the same as in Figure \ref{fig:mcmc_i_gi_rz}.
}
\label{fig:mcmc_w1}
\end{figure*}

Adopting this multi-linear model for $\Upsilon^*_{W1}$, we follow the same MCMC procedure to map the posterior distributions of the BTFR parameters using $W1$ band photometry. Results are summarized in Figure~\ref{fig:mcmc_w1}. Here, only Slope and ZP are adopted as free parameters, because the parameters describing $\Upsilon^*_{W1}$ have been already found by the analysis of the galaxies that are in common between our optical and infrared sub-samples. Fixing the $\Upsilon^*_{W1}$ parameters guarantees that the estimated stellar mass to light ratios are in agreement with those estimated using the optical data and improves the consistency between the BTFR parameters we find in different scenarios.

The parameters of the fitted BTFR with the $W1$ band photometry are included in Table~\ref{tab:mcmc-params}. The infrared correlation is illustrated in the bottom right panel of Figure~\ref{fig:BTFR-nonCalibrated}. The WISE infrared and SDSS optical versions are consistent with each other thanks to the coupling imposed through Equation~\ref{eq:op_ir_coupled}.


\section{Zero Point Calibration}
\label{sec:zp}

\begin{figure*}
\centering
\includegraphics[width=0.46\linewidth]{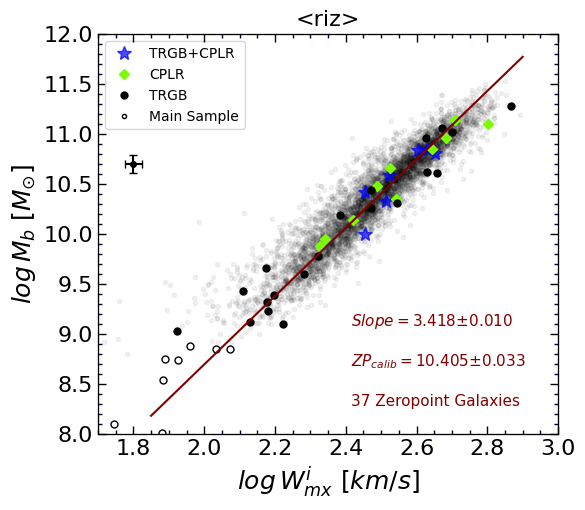}
\includegraphics[width=0.46\linewidth]{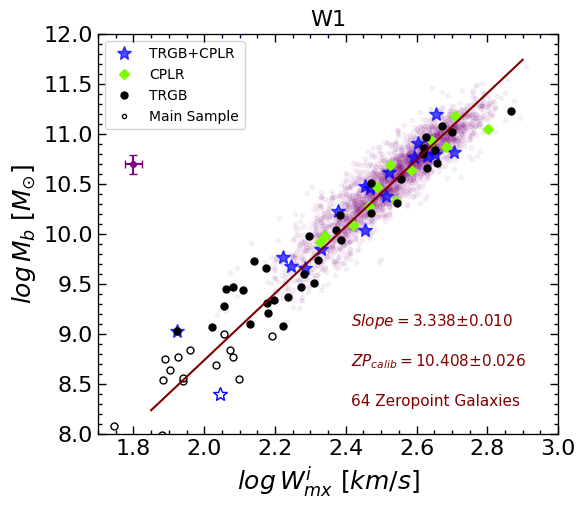} 
\caption{BTFR with SDSS optical photometry (left) and WISE infrared photometry (right) after compensation for the \hi selection bias. Zero point calibrators are represented by filled symbols where $\widetilde{M}_b > 10^9$ $M_\odot$, and by open symbols otherwise. The legend in each panel identifies the components of the calibration in each instance.  The main sample galaxies are shifted vertically to be consistent with the calibrated zero point. The typical uncertainty of the scattered points is displayed on the top left corner.
}
\label{fig:BTFRZP}
\end{figure*}



To this point, the construction of the BTFR has been based on distances assuming $H_0=75$~\kmsMpc.  Now, the scaling of (pseudo) baryonic mass is shifted to match values for galaxies with measured distances from either the Cepheid Period-Luminosity Relation (CPLR) or the magnitudes of stars at the Tip of the Red Giant Branch (TRGB).  The galaxies used here and their measured distances are the same as those providing the absolute calibration in \citet{2020ApJ...896....3K, 2020ApJ...902..145K}.
To summarize, in the case of the CPLR the dominant sources of distance information are \citet{2001ApJ...553...47F} and \citet{2016ApJ...826...56R} rescaled slightly as described by \citet{2019ApJ...876...85R} for consistency with the LMC detached eclipsing binary distance by \citet{2019Natur.567..200P} and the NGC\,4258 maser distance by \citet{2013ApJ...775...13H}.  In the case of the TRGB the source in all cases is the compilation maintained in the Color-Magnitude Diagrams/TRGB file in the Extragalactic Distance Database (https://edd.ifa.hawaii.edu) that was described most recently by \citet{2021AJ....162...80A}, with the zero-point calibration established by \citet{2007ApJ...661..815R}.  The distances given there are demonstrated by \citet{2021arXiv210800007A} to be compatible within uncertainties with those discussed by \citet{2021ApJ...919...16F}.  Any CPLR or TRGB measurement available from these sources is used if it is associated with a spiral galaxy that meets BTFR criteria.  At optical bands these give 39 calibrators: 19 CPLR and 29 TRGB with 9 in common.  In the infrared there are 64 calibrators: 29 CPLR and 52 TRGB with 17 in common.  There are more calibrators available in the infrared because of the full sky coverage.

Figure~\ref{fig:BTFRZP} presents the BTFR relations shifted to match the constraints of the calibrator systems.  In the right panel, baryonic masses are determined from WISE $W1$ magnitudes and gas masses.  This plot corresponds to the bottom right panel of Figure~\ref{fig:BTFR-nonCalibrated} except now zero point calibrators are superimposed and the underlying sample is slightly shifted vertically to match the calibrators.  The panel to the left is similar but the BTFR constitutes an average of the SDSS optical material as in the lower left panel of Figure~\ref{fig:BTFR-nonCalibrated}.

The infrared and optical calibrations seen in Figure~\ref{fig:BTFRZP} are in statistical agreement in both slopes and zero points.  
Concerning the slopes, in the absence of dissipative or feedback  effects, the response of motions to mass in a virialized structure in a $\Lambda$CDM cosmology should have the dependence $M \propto V^3$ \citep{1997gsr..proc....3W, 2010AJ....140..663G}.  The interplay between baryons and dark matter during the process of galaxy formation evidently causes the relation between mass and motion to steepen.  Constraints on the BTFR slope provide constraints on semi-analytic models of galaxy formation although details depend on interpretations of the observed parameters, in particular line widths as a proxy for motions \citep{2016ApJ...832...11B, 2019MNRAS.484.3267L}.  Slopes in the logarithmic relation between baryonic mass and rotation rates in the literature have tended to lie around 3.9 if rotation curves are resolved and 3.3 if the rotation estimator is a global profile.  Our results are consistent with the latter, as we find a slope of 3.34 at the infrared W1 band and 3.42 with the combined $r,i,z$ optical bands.

Since optical and infrared $ZP$ and $Slope$ are slightly different, measured distances will be slightly deviant between the two (evaluated in detail further along). 
  It is to be recalled that the sky coverage of the WISE infrared and SDSS optical components of our sample are overlapping but not contiguous.  Streaming motions could cause small offsets.  Consequently, we retain the separate WISE and SDSS calibrations for the measurements of individual galaxy distances.  If both are available, a straight average of the two is taken.
Our formulation of the BTFR is poorly defined at masses below $10^9 \Msun$ so galaxies less massive than this limit are excluded.

\section{Calculation of Distances}
\label{sec:distMeasure}

Input observables are SDSS optical and/or WISE infrared apparent magnitudes, \hi fluxes ($F_{21}$), \hi linewidths (${\rm log} W^i_{mx}$) and radial velocities that are used to make the gas bias adjustments.
The calculation of the distance modulus of each galaxy incorporates the optimized parameters for the construction of the BTFR given in Table~\ref{tab:mcmc-params} and ${\rm log} \Upsilon^*_{\lambda}$ as described in $\S \ref{sec:optimizeOP}$ and $\S \ref{sec:optimizeIR}$.  The calculations will be described in a reverse order compared to what we presented earlier. 

Given an \hi linewidth, the BTFR allows calculation of the galaxy quasi baryonic mass 
\begin{equation}
    \widetilde{M}_b = 10^{(Slope ({\rm log}W^i_{mx}-2.5) + ZP)}.
\label{eq:INVcBTFR}
\end{equation}
On the other hand, if the galaxy distance is given, one can alternatively determine the baryonic mass by calculating its components, gas and stellar mass, through Equation \ref{eq:Mb}. We denote this distance dependent estimated mass by $\widetilde{M}_b(d)$. The dependency of the gas mass estimation with distance is obvious in Equation \ref{eq:MHI}. The stellar mass is derived using $M_*=\Upsilon^*_{\lambda}L_{\lambda}$, where $\Upsilon^*_{\lambda}$ is derived from color terms that are distance independent (Equation \ref{eq:colorterm} and Figure \ref{fig:ms_w1_w2}). However, $L_{\lambda}$, the galaxy absolute luminosity, is derived from its apparent luminosity and a specified distance. Accordingly, for each galaxy there is only one distance that results in the same $\widetilde{M}_b(d)$ as that given by the BTFR, $\widetilde{M}_b$. For each galaxy, we numerically examine all distances in the range of $0.5$ and $500$ Mpc with step sizes $0.5$~Mpc. To solve for the distance, we interpolate between two adjacent distances, $d_1$ and $d_2$, for which $\widetilde{M}_b(d)-\widetilde{M}_b$ takes opposite signs.

In the vocabulary of Eq.~\ref{eq:d_est}, $\widetilde{M}_b$ encodes the expectation baryonic mass along the fiducial BTFR relation at its proper distance while $\widetilde{M}_b(d)$ is the observed baryonic mass at the distance required to place the target on the fiducial relation.  The shift of $\widetilde{M}_b(d)$ from $\widetilde{M}_b(d_{H_0})$ (where $H_0 \simeq 75$~\kmsMpc\ as discussed in \S \ref{sec:H0}) varies with mass as $d^2$.

It is to be recalled that the BTFR measurements are in the logarithm of mass and we are looking for offsets from fiducial with roughly symmetric scatter in distance moduli, proportional to the log of distances.  The consequence is an asymmetry in the scatter of distance estimates about unbiased values, which propagates into bias in estimates of deviations from Hubble flow: peculiar velocities.  Translation of our raw distances (properly, distance moduli) into peculiar velocity constructs is beyond the scope of this study.  Various procedures have been developed to infer peculiar velocities from estimated distance moduli with large uncertainties \citep{2015MNRAS.450.1868W, 2019MNRAS.488.5438G, 2021MNRAS.505.3380H,2021MNRAS.507.1557L}.

\begin{figure}
\centering
\includegraphics[width=0.95\linewidth]{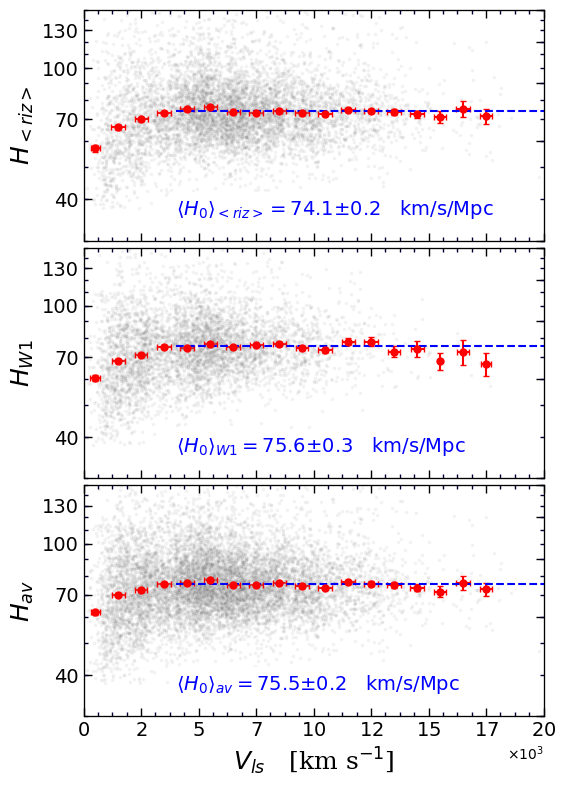} 
\caption{Hubble parameter as a function of radial velocity. Galaxies are represented by black points.
Blue horizontal line shows the log average of the Hubble parameter of galaxies beyond 4000~\kms.  Red points display the average of the Hubble parameter within velocity intervals of 1000~\kms, with open points representing the the averages at intervals less than 4000~\kms. Top and middle panels correspond to distances that are obtained from the BTFRs in the left and right panels of Figure \ref{fig:BTFRZP}, respectively. In the bottom panel, the optical and infrared distances are averaged where they are both available.
Zero points uncertainties are not incorporated in the reported errors in legends.
\label{fig:H0}}
\end{figure}

\begin{figure*}
\centering
\includegraphics[width=0.46\linewidth]{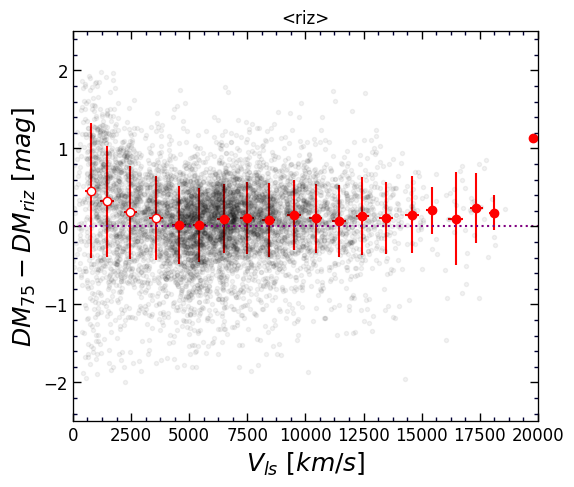} 
\includegraphics[width=0.46\linewidth]{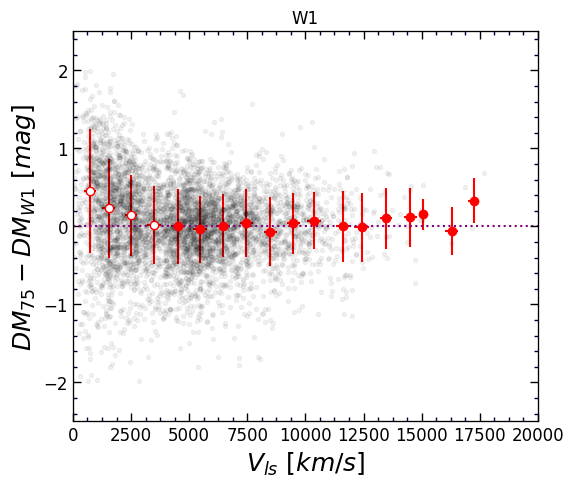} 
\includegraphics[width=0.46\linewidth]{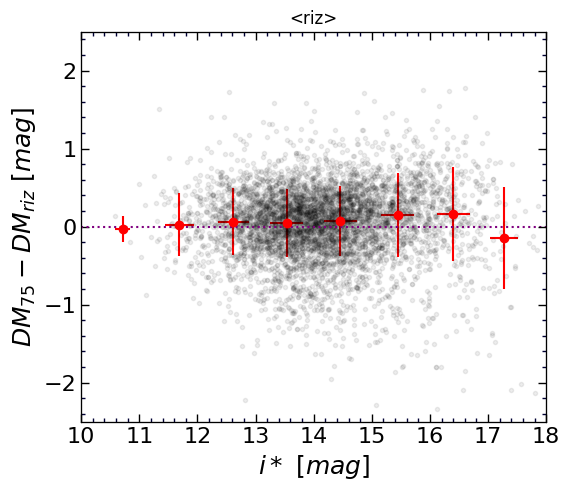} 
\includegraphics[width=0.46\linewidth]{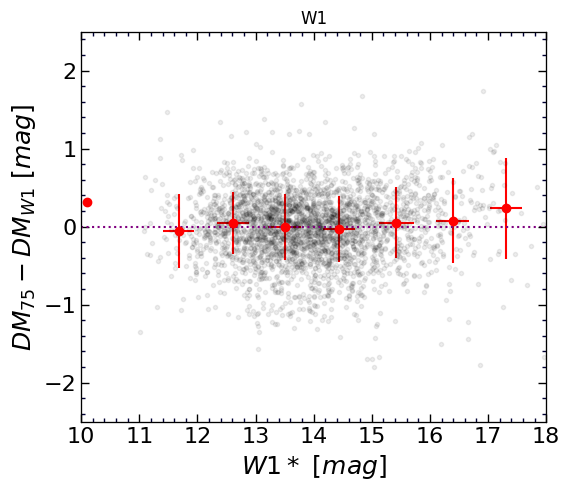} 
\caption{{\bf Top:} Deviations from Hubble law distances as a function of systemic velocity. SDSS optical photometry case at left and WISE infrared photometry case at right. Open symbols represent galaxies with $V_{ls}<4000$ \kms, where Hubble law distances are not reliable distance proxies.
{\bf Bottom:} Demonstration that there are no systematic effects as functions of either SDSS optical photometry (left) or WISE infrared photometry (right) for galaxies with $V_{ls}>4000$ \kms.
}
\label{fig:2Vdeld}
\end{figure*}
\subsection{Uncertainties}
\label{sec:uncertaintyMeasure}

In the previous section, we described how a galaxy distance is calculated for a given set of observables and model parameters. Here, prior to calculating the uncertainty of the measured distances, we study the contribution of each error term in the scatter of data points about the BTFR and subsequent distance measurement uncertainties.  The 1$\sigma$ scatter of our sample galaxies is $0.18$ dex about the direction of baryonic mass in the BTFR diagram. 
This scatter is roughly equivalent to an uncertainty of $0.45$ mag in the distance modulus measurements. We want to evaluate the contribution of different sources of uncertainty to the ensemble value. 

In our study, uncertainties have mainly four sources: (1) uncertainties associated to the model parameters that describe correlations between different quantities, i.e. the errors on the fitted coefficients, (2) errors that are related to the scatter of the model residuals that basically monitors how well the constructed model performs, (3) uncertainties of the observables that are used in the model, and (4) the unknown uncertainties that are related to the underlying physical processes that govern the dynamic and composition of our sample galaxies. 

To estimate the effect of each error term, first we generate an ensemble of 10,000 galaxies by randomly drawing from the original sample. In this process, a galaxy may appear multiple times in the ensemble, however its parameters would be dispersed differently.
For each galaxy in the ensemble, we adopt the corresponding baryonic mass and distance from the original sample. Assuming that the simulated galaxies completely obey the BTFR gives us a way to compute their linewidths. By introducing uncertainties of different natures to the parameters of the ensemble galaxies, we measure how galaxies are dispersed along the vertical axis of BTFR which is then translated to the error in the distance modulus by $d\/DM = 2.5\/ d {\rm log}M_b$.

Below, we calculate the contribution of various uncertainties in the total scatter of the BTFR. For the sake of discussion, the BTFR and other relevant parameters are taken from our $riz$ model. The error budget is roughly the same if we consider other BTFRs at different wavebands. In this analysis the values and uncertainties of the \hi flux ($F_{21}$) and linewidth (${\rm log}W^i_{mx}$) are taken from Table 1 of \citet{2020ApJ...902..145K} and given again in \S \ref{sec:table} Table~\ref{tab:catalog}. The uncertainties on the apparent magnitudes at SDSS optical ($r$, $i$, and $z$) and WISE infrared ($W_1$ and $W_2$) bands are less than $0.05$ mag, the value we conservatively adopt in this analysis. 

\begin{itemize}
    
    \item Uncertainties in the calculated stellar mass to light ratios that have three sources. One originates from the errors in the coefficients $\alpha_{\lambda}$ and $\beta_{\lambda}$ in Equation \ref{eq:colorterm}. The other is related to the uncertainty of the $g-i$ color, which is $\sim 0.07$ mag considering that the uncertainties on the apparent magnitudes at SDSS optical ($r$, $i$, and $z$) and WISE infrared ($W_1$ and $W_2$) bands are less than $0.05$ mag. Given the rms scatter of the residuals of the $\Upsilon^*_{\lambda}$ models, we associate an uncertainty of $0.1$ dex on the logarithm of the calculated stellar mass to light ratios, $\rm{log} \Upsilon^*_{\lambda}$ $[\Msun/\Lsun]$ (see Figures \ref{fig:linkopir} and \ref{fig:ms_w1_w2}). This level of uncertainty is equivalent to an average of $\sim 20\%$ error on the derived $\Upsilon^*_{\lambda}$. Dispersing the stellar mass to light ratios of the ensemble test galaxies and recalculating their total baryonic mass, results in an scatter of $0.08$ dex on the ${\rm log} \widetilde{M}_b$ which implies an uncertainty of $0.2$ mag on the distance modulus.
    
    \item Uncertainty of the gas mass. Assuming a Gaussian error of $0.1$ on $K_g$ and in calculation of the baryonic mass from Equation~\ref{eq:Mb} generates a scatter of 0.01 dex on ${\rm log} \widetilde{M}_b$ which is equivalent to the moduli uncertainty of $0.02$ mag. Another error factor in the calculation of the gas mass is the error of the measured \hi flux. Dispersing the data points based on the error of the \hi flux generates a dispersion of less than $0.01$ dex along ${\rm log} \widetilde{M}_b$, equivalent to an error of $0.02$ mag in distance modulus.  The combined error of 0.03 is minor.

    \item The error of the velocity multiplier $C_{adj}$ in Equation \ref{MHIcorrection} and the scatter of residuals about the linear correlation shown in the middle panel of Figure \ref{fig:delHI}. This scatter has been modeled by the dashed line in the bottom panel of Figure \ref{fig:delHI}. Considering these two types of errors introduces an scatter of $0.08$ dex along ${\rm log} \widetilde{M}_b$, an error of $0.2$ mag on distance modulus.
    
    \item Uncertainty of the \hi line width, ${\rm log}W^i_{mx}$. We randomly disperse the linewidth of the simulated galaxies according to the error on the ${\rm log}W^i_{mx}$ parameter which already includes the uncertainty of the linewidths and inclinations measurements. A dispersion of $0.1$ dex on ${\rm log} \widetilde{M}_b$ is implied which is translated to $\sim 0.25$ mag error on the distance modulus.
    
    \item Adding all mentioned known uncertainties we end up with a scatter of $0.15$ dex in ${\rm log} \widetilde{M}_b$ which is smaller than the 0.18 dex that we observe overall. Following the quadratic rule of addition of Gaussian errors, this unexplained scatter is equivalent to $0.1$ dex in ${\rm log} \widetilde{M}_b$ that translates to $23\%$ uncertainty on the calculated baryonic mass from the BTFR correlation. We attribute this residual to the intrinsic uncertainty of the BTFR correlation which could originate from unknown physical processes such as variations between the distributions of baryonic and dark matter.  We include this unknown uncertainty in  our analysis to get more reasonable errors on the measured distances.
    
\end{itemize}

Now that we have estimated the sources and ensemble contributions of different uncertainties in the final measurements, we can focus on each galaxy individually. In practice, we measure the distance of each galaxy using an ensemble of 10,000 sets of parameters that are generated by randomly dispersing the input observables based on their reported Gaussian errors. In addition to accounting for the uncertainties of the galaxy observables, each measurement adheres to the BTFR and $\Upsilon^*_{\lambda}$ models whose parameters are drawn from either their posterior distributions or the covariance matrices that are generated in the maximum likelihood fitting processes. The scatter about the fitted correlations is considered in the same process as explained above. In the end, we add an additional uncertainty of $0.1$ dex on ${\rm log} \widetilde{M}_b$ when measuring distances of the ensemble galaxies. For each galaxy, we perform 10,000 measurements using the simulated parameters and we report the median of all deduced distances as the final galaxy distance and adopt the $1\sigma$ scatter as the statistical error.

\begin{table*}
\centering
\caption{The BTFR zero points and corresponding $H_0$ values for different calibrator samples
\label{tab:calibrators_zp_H0}}
\begin{tabular}{c | c c | c c c}
\hline \hline
Calibrator & $ZP_{<riz>}$ & $ZP_{<W1>}$ & $\langle H_0\rangle_{<riz>}$ & $\langle H_0\rangle_{<W1>}$ & $\langle H_0\rangle_{av}$ \\
Sample & \multicolumn{2}{c}{\tt BTFR Zero Point}  & \multicolumn{3}{c}{\tt [\kmsMpc]} \\

\hline
CPLR & $10.401\pm0.037$ & $10.406\pm0.030$ & $74.5\pm3.2$ & $75.8\pm2.6$ & $75.8\pm2.8$ \\
TRGB & $10.406\pm0.041$ & $10.409\pm0.030$ & $74.1\pm3.6$ & $75.6\pm2.6$ & $75.5\pm3.0$ \\
\hline
CPLR+TRGB & $10.405\pm0.033$ & $10.408\pm0.026$ & $74.1\pm2.8$ & $75.6\pm2.3$ & $75.5\pm2.5$ \\
\hline
\end{tabular}
\end{table*}

\section{The Hubble Constant}
\label{sec:H0}

The zero point fits enable a determination of the Hubble Constant consistent with this data.  The arbitrary value of $H_0=75$~\kmsMpc\ was taken to study relative properties of the BTFR in $\S2$ because our studies with the same data \citep{2020ApJ...896....3K, 2020ApJ...902..145K} favored approximately this value.  
The zero point fits shown in Figure~\ref{fig:BTFRZP} are consistent with this initial choice of $H_0$ to within one Hubble unit.  

Individual galaxy Hubble parameter values following Eq.~\ref{eq:hubbleparameter} with 
distances derived from deviations from the separate $W1$ and $<riz>$ relations in Figure~\ref{fig:BTFRZP} are displayed as a function of systemic velocity in Figure~\ref{fig:H0}.  The fits shown by dashed blue lines are averaged over velocities greater than 4,000~\kms\ where deviations due to peculiar velocities are expected to be small. The Hubble parameters are averaged in logarithmic space, where the uncertainties are approximately Gaussian, and are weighted using the moduli uncertainties discussed in the previous section.
The rms scatter corresponds to a scatter of 22\% in distances.
The uncertainties quoted in the figure legends are statistical given the absolute calibration.  The statistical uncertainties associated with the zero point calibration are larger: $\Delta H_0=\pm2.3$ and $\pm2.8$ at $W1$ and $<riz>$ respectively.  

As shown in Figure~\ref{fig:H0}, we find
$H_0^{W1}=75.6\pm2.3$~\kmsMpc\ and $H_0^{<riz>}=74.1\pm2.8$~\kmsMpc\ where the errors are the quadrature sum of the scatter seen in Figure~\ref{fig:H0} and the zero point scale uncertainties of Figure~\ref{fig:BTFRZP}. The errors do not include uncertainties in the Cepheid and TRGB absolute scales.  Combining the optical and infrared measures we find $H_0^{av}=75.5\pm2.5$~\kmsMpc. This value differs from a straight average of the optical and infrared values because the sky coverages in the separate bands are different.

We pursue the same methodology for determining the BTFR zero point and Hubble constant using each set of calibrator separately. Table \ref{tab:calibrators_zp_H0} lists the BTFR zero points and their associated Hubble constants. The deduced values for different calibrator samples are in agreement considering the 1$\sigma$ uncertainties. The reported distances in this study are constructed based on the zero points we found for the $TRGB+CPLR$ sample.

\begin{figure}
\centering
\includegraphics[width=0.95\linewidth]{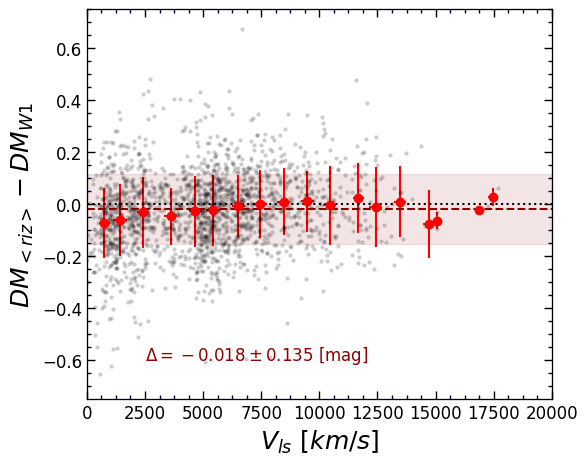}
\caption{Comparison of distance moduli derived alternatively from optical and infrared photometry as a function of systemic velocity.
\label{fig:Vdif}}
\end{figure}

\begin{figure}
\centering
\includegraphics[width=0.99\linewidth]{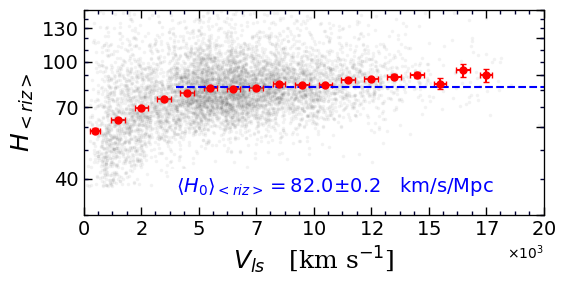} 
\caption{
NO \hi CORRECTION. The treatment leading to this plot is identical to that giving the middle panel of Fig.~\ref{fig:H0} except the adjustment for the \hi flux bias has {\it not} been applied.
\label{fig:nocorrection}}
\end{figure}
\section{Tests}
\label{sec:tests}

There are necessary, though not sufficient, tests that a sample must pass.  For example, at systemic velocities greater than $\sim4,000$~\kms\ deviant velocities are expected to be small compared with Hubble expansion velocities.  The log of distance estimates should scatter around the Hubble expansion expectation.\footnote{
There is a long history of the discussion of biases that can arise with distance measurements suffering substantial uncertainties. An acute example is a potential sample selection bias, discussed in the context of luminosity$-$linewidth measurements by, among others, \citet{1994ApJ...430...13S, 1999AJ....117..157S} and \citet{1994ApJS...92....1W}. 
\citet{2000ApJ...533..744T} provide a substantial review.  In the current application, the dominant selection effect is imposed by the \hi flux limit.  Our procedure for negating the bias that could arise is tested by simulations as discussed in \S \ref{sec:mock-test}.  A characteristic signature of bias is a variation of Hubble parameter with redshift or another independent variable such as apparent magnitude or inclination.
}
The top panels of Figure~\ref{fig:2Vdeld} illustrates the constancy of the ratio, observed distances to Hubble flow expectation, in bins of observed velocity.  Results are equally good with infrared and optical photometry (results with $r$ and $z$ are almost indistinguishable from those at $i$). Figure~\ref{fig:Vdif} provides a direct comparison between the infrared and $<riz>$ averaged optical material when all are available.  There is a statistically insignificant offset of $0.8\%$ in average distances. The necessary test is passed.

Figure~\ref{fig:nocorrection} draws on data that are in all respects the same as that shown in the bottom panel of Figure~\ref{fig:H0} except the baryonic masses have {\it not} received the \hi flux bias adjustment.  The problem without the adjustment is clear.  Hubble parameter values drift higher with redshift, just as was seen with the mock sample in the upper panel of Figure~\ref{fig:mockHV}.  For galaxies with the same apparent magnitudes at the same redshifts, those with greater \hi flux (larger relative baryonic mass) are retained while those with lesser \hi flux are lost.  These retained galaxies tend to lie to the left of the fiducial BTFR, hence are given distances less than a case on the fiducial relation, leading to increased $H_0$ values.  Without the \hi flux bias adjustment, any estimate of $H_0$ is compromised.

Another test is to evaluate the run of distance estimates normalized to Hubble flow expectations as a function of apparent infrared and optical magnitudes.  Again, scatter about a constant ratio is expected.  It has been demonstrated that measured distances in the compilation of \citet{2020ApJ...902..145K} failed this test.  That analysis did not adequately account for the curvature in the TFR and motivated this study.  As seen in the bottom panels of Figure~\ref{fig:2Vdeld}, the distances determined here pass this test with both infrared and optical photometry.

\begin{figure*}
\centering
\includegraphics[width=0.46\linewidth]{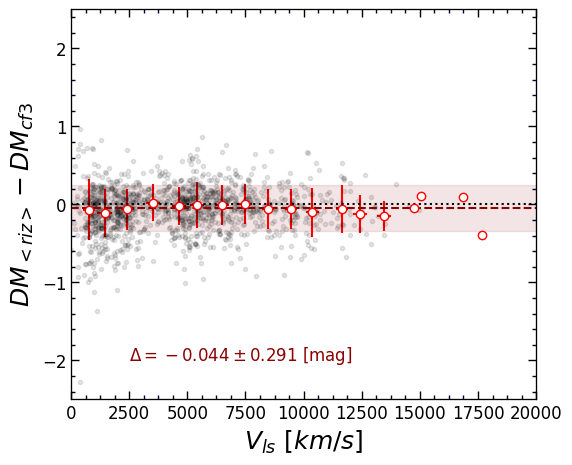} 
\includegraphics[width=0.46\linewidth]{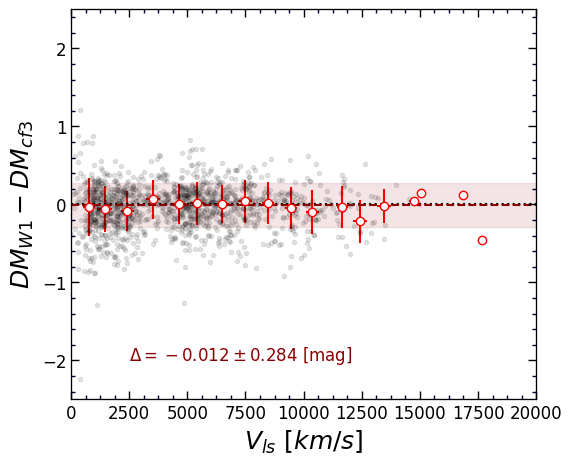} 
\caption{Comparison of distance moduli reported in {\it Cosmicflows-3} with BTFR distance moduli (left: SDSS optical photometry; right: WISE infrared photometry).  There are small zero point differences.
\label{fig:2CF3}}
\end{figure*}

A further test is to compare the new BTFR distances with TFR distances reported in {\it Cosmicflows-3} \citep{2016AJ....152...50T}.  It is seen in Figure~\ref{fig:2CF3} that there is constancy in the velocity bin averages with both infrared and the $<riz>$ optical photometry.  There are zero point offsets but we leave reconciliation to the ensemble {\it Cosmicflows-4} compilation.


\section{Table of Distances}
\label{sec:table}

The distances of 10154 galaxies in our sample and their observed/inferred information that are used in this study are cataloged in Table \ref{tab:catalog}.  See also the table {\it CF4 BTF-distances} maintained and updated in the {\it Extragalactic Distance Database} (https://edd.ifa.hawaii.edu).
Columns are 

(1) The ID number of the galaxy in the Principal Galaxy Catalog (PGC; \url{http://leda.univ-lyon1.fr/}). 

(2) Flag: 1 = accepted; 0 = rejected as either $>3.5 \sigma$ deviant or below the mass limit of $10^9 \Msun$ or if the modulus uncertainty is larger than $0.8$ mag.

(3) The average of distance moduli measured using optical and infrared data. The reported uncertainty is the maximum adjusted error of the optical and infrared moduli.

(4) The measured distance modulus from the averaged optical BTFR, $<riz>$ (refer to the left panel of Figure \ref{fig:BTFRZP}). The statistical error on the measured moduli using the optical data derived from the Gaussian propagation of uncertainties in the associated measured quantities (refer to \S \ref{{sec:uncertaintyMeasure}} for detailed discussion).


(5) The measured distance modulus from the $W1$ BTFR (refer to the right panel of Figure \ref{fig:BTFRZP}), and their uncertainties. 


(6) The PGC ID of the dominant galaxy in the parent group \citep{2017ApJ...843...16K}.

(7) The parent group ID in the 2MASS group catalog \citep{2015AJ....149..171T}.

(8) The logarithm of the average stellar mass calculated based on SDSS photometry data at the $r$, $i$ and $z$ bands.

(9) The logarithm of the stellar mass calculated based on $W1$ band photometry data assuming the mass to light ratio, $\Upsilon^*_{W1}$, defined by the fit in Figure~\ref{fig:ms_w1_w2}. 

(10) The logarithm of the galaxy gas mass calculated using $1.33M_{HI}$, where $M_{HI}$ is the atomic hydrogen mass.

(12) The logarithm of the total baryonic mass of a galaxy in stellar and gas forms.

(13) Heliocentric velocity from \hi observations.

(14) Radial velocity relative to the Local Sheet \citep{2008ApJ...676..184T}. 

(15) Radial velocity in the rest frame of the cosmic microwave background. 

(16) The cosmological correction factor 
$f_j=1+1/2(1-q_0)z_j-1/6(1-q_0-3q_0^2+j_0)z_j^2$ where $z_j$ is the redshift of the galaxy, $j_o\simeq1$, and $q_0=0.5(\Omega_m-2\Omega_{\Lambda})=-0.595$ \citep{2004CQGra..21.2603V} assuming $\Omega_m=0.27$ and $\Omega_{\Lambda}=0.73$ .

(17) The logarithm of the inclination-corrected \hi line width, calculated from $W^i_{mx}=W_{mx}/\sin(i)$, where $i$ is the inclination angle presented in column (18).  The line width parameter $W_{mx}$ is derived from a measure of the 50 percentile width of an observed \hi profile, adjusted for instrumental resolution and redshift stretch effects, and statistically descriptive of the peak-to-peak maximum rotation velocity of a galaxy \citep{2009AJ....138.1938C, 2011MNRAS.414.2005C}, with uncertainty.   

(18) \hi 21cm flux, with uncertainty.

(19) The inclination of the galaxy in degrees, with uncertainty. 

(19-24) The {\it SDSS g,r,i,z} and {WISE \it W1,W2} magnitudes in the AB system, corrected for Milky Way obscuration, redshift $k-$correction, and the host dust attenuation. Please refer to \S 2.2.2 of \citet{2020ApJ...902..145K} for further details. The uncertainty of observed magnitudes is $0.05$ mag.

(23-24) The manually assigned quality for the photometry of SDSS and WISE images. The quality grade ranges from $0$ for the poorest quality (or missing data) to $5$ for the best quality. 

(27-32) Galaxy coordinates in equatorial, galactic and supergalactic reference frames.

(33) Number of galaxies in the parent group identified in \citet{2017ApJ...843...16K}.

(34) Number of galaxies in the parent group identified in \citet{2015AJ....149..171T}.

(35-36) The coordinates of parent group in the supergalactic frame of reference.

(37-39) The average velocity of the parent group in the heliocentric, Local Sheet, and CMB rest frames. 

(40) The cosmological correction factor described in column (17), calculated using the redshift of the parent group.

\clearpage
\begin{landscape}

\begin{table}
\centering
\caption{Data Catalog$^{\dag}$}
\label{tab:catalog}
\setlength{\tabcolsep}{3pt}
\begin{tabular}{r ccccccrrccccccccccc}
\hline \hline
PGC & 
Flag &
$DM_{av}$ & 
$DM_{riz}$ & 
$DM_{W1}$ & 
1PGC & 
Nest & 
$\log(M_*)^a$ & 
$\log(M_*)^b$ & 
$\log(M_g)^c$ & 
$\log(M_b)^d$ & 
$V_{h}$ & 
$V_{LS}$ & 
$V_{CMB}$ &  
$f$ &
$\log (W^i_{mx)}$ & 
$F_{21}$ & $Inc.$ 
\\
 & 
 &
(mag) & 
(mag) & 
(mag) & 
 & 
 &
$[M_\odot]$ & 
$[M_\odot]$ & 
$[M_\odot]$ & 
$[M_\odot]$ &
(km/s) & 
(km/s) & 
(km/s) & 
 & 
 & (deg)
Jy$\cdot$\kms 
\\
(1) & (2) & (3) & 
(4) & (5) & (6) & 
(7) & (8) & (9) & 
(10) & (11) & (12) & 
(13) & (14) &
(15) & (16) & (17) &
(18)  \\
\hline
2 & 1 & 34.51$\pm$0.46 &  & 34.49$\pm$0.46 & 73150 & 200275 &  & 11.154 & 9.910 & 11.178 & 5004 & 5296 & 4726 & 1.013 & 2.744$\pm$0.029 & 4.73$\pm$0.81 & 52$\pm$4 \\
4 & 1 & 33.49$\pm$0.49 & 33.43$\pm$0.48 & 33.72$\pm$0.49 & 120 & 202766 & 9.394 & 9.266 & 9.377 & 9.655 & 4458 & 4706 & 4109 & 1.011 & 2.189$\pm$0.014 & 1.91$\pm$0.07 & 85$\pm$2 \\
12 & 1 & 34.97$\pm$0.41 &  & 34.99$\pm$0.41 & 12 & 210177 &  & 10.647 & 9.939 & 10.724 & 6548 & 6685 & 6195 & 1.016 & 2.606$\pm$0.021 & 3.40$\pm$0.58 & 82$\pm$3 \\
16 & 1 & 34.63$\pm$0.45 & 34.62$\pm$0.45 & 34.66$\pm$0.42 & 16 & 211419 & 10.373 & 10.371 & 9.362 & 10.413 & 5667 & 5809 & 5312 & 1.014 & 2.515$\pm$0.025 & 1.19$\pm$0.20 & 65$\pm$4 \\
55 & 1 & 33.83$\pm$0.60 & 33.81$\pm$0.59 & 33.95$\pm$0.60 & 55 &  & 9.344 & 9.276 & 9.791 & 9.915 & 4779 & 5052 & 4454 & 1.012 & 2.260$\pm$0.025 & 4.30$\pm$0.32 & 80$\pm$3 \\
68 & 1 & 34.71$\pm$0.56 & 34.82$\pm$0.56 & 34.71$\pm$0.53 & 68 &  & 10.067 & 10.148 & 9.738 & 10.262 & 7664 & 7740 & 7338 & 1.019 & 2.390$\pm$0.043 & 1.61$\pm$0.27 & 57$\pm$4 \\
70 & 1 & 35.18$\pm$0.41 & 35.32$\pm$0.41 & 35.15$\pm$0.39 & 70 & 209949 & 10.603 & 10.691 & 10.369 & 10.831 & 6800 & 7040 & 6447 & 1.017 & 2.636$\pm$0.005 & 8.33$\pm$0.12 & 90$\pm$1 \\
76 & 1 & 34.76$\pm$0.41 & 34.72$\pm$0.41 & 34.83$\pm$0.39 & 76 & 209247 & 10.873 & 10.835 & 10.109 & 10.926 & 6920 & 7183 & 6583 & 1.017 & 2.624$\pm$0.013 & 4.42$\pm$0.11 & 68$\pm$4 \\
92 & 1 & 32.82$\pm$0.50 & 32.89$\pm$0.50 &  & 92 &  & 9.612 &  & 9.818 & 10.028 & 5376 & 5592 & 5015 & 1.013 & 2.165$\pm$0.016 & 3.78$\pm$0.08 & 80$\pm$3 \\
94 & 1 & 33.83$\pm$0.58 &  & 33.86$\pm$0.58 & 94 &  &  & 9.374 & 9.593 & 9.799 & 4098 & 4367 & 3995 & 1.011 & 2.274$\pm$0.032 & 3.61$\pm$0.62 & 90$\pm$1 \\
\dots \\
\hline
\end{tabular}
\end{table}


\begin{table}
\centering
\addtocounter{table}{-1}
\caption{Data Catalog (continued)$^{\dag}$}
\label{tab:catalog}
\setlength{\tabcolsep}{3pt}
\begin{tabular}{c cccccccccrrrrrrccrrcccc}
\hline \hline
PGC & 
$g^*$ & $r^*$ &  
$i^*$ &  $z^*$ & $W1^*$ &  $W2^*$ & 
$Q_{S}$ & $Q_{W}$  &
RA & Dec & Glon & Glat & SGL & SGB &
$N_{kt}$ & $N_{15}$ &
SGL$^{(g)}$ & SGB$^{(g)}$ & $V_{h}^{(g)}$ & $V_{LS}^{(g)}$ & $V_{CMB}^{(g)}$ & $f_g$ 
\\
 & 
(mag) & (mag) & (mag) & 
(mag) & (mag) & (mag) &
 &  & 
(deg) & (deg) & (deg) & 
(deg) & (deg) & (deg) &
 &  & 
(deg) & (deg) & (km/s) & 
(km/s) & (km/s) &
\\
(1) &  (19) & (20) & (21) &
(22) & (23) & (24) &
(25) & (26) & (27) & 
(28) & (29) & (30) & 
(31) & (32) & (33) & 
(34) & (35) & (36) & 
(37) & (38) & (39) &
(40)  \\
\hline
2 &  &  &  &  & 11.89 & 12.52 &  & 5 & 0.0005 & 47.2745 & 113.9553 & -14.6992 & 341.6440 & 20.7388 & 0 & 7 & 341.4922 & 20.7395 & 5194 & 5486 & 4916 & 1.013 \\
4 & 15.59 & 15.33 & 15.12 & 15.07 & 16.07 & 16.44 & 5 & 4 & 0.0010 & 23.0876 & 107.8322 & -38.2729 & 316.0587 & 18.4514 & 0 & 0 & 316.0587 & 18.4514 & 4458 & 4706 & 4109 & 1.011 \\
12 &  &  &  &  & 13.59 & 14.23 &  & 5 & 0.0024 & -6.3739 & 90.1920 & -65.9300 & 286.4249 & 11.3511 & 0 & 1 & 286.4249 & 11.3510 & 6532 & 6669 & 6179 & 1.016 \\
16 & 14.36 & 13.86 & 13.58 & 13.40 & 13.99 & 14.70 & 5 & 5 & 0.0031 & -5.1587 & 91.6005 & -64.8656 & 287.6119 & 11.7030 & 0 & 1 & 287.6120 & 11.7030 & 5709 & 5851 & 5354 & 1.014 \\
55 & 15.53 & 15.32 & 15.22 & 15.17 & 16.20 & 16.63 & 5 & 4 & 0.0104 & 33.6009 & 110.9496 & -28.0857 & 327.0996 & 19.7763 & 0 & 0 & 327.0996 & 19.7763 & 4779 & 5052 & 4454 & 1.012 \\
68 & 14.93 & 14.61 & 14.52 & 14.34 & 15.03 & 15.54 & 5 & 5 & 0.0154 & -18.9589 & 65.4189 & -75.8101 & 274.3903 & 7.1770 & 0 & 0 & 274.3903 & 7.1770 & 7664 & 7739 & 7338 & 1.019 \\
70 & 13.87 & 13.46 & 13.24 & 13.02 & 13.54 & 14.11 & 5 & 5 & 0.0156 & 20.3380 & 107.1780 & -40.9837 & 313.2487 & 17.7662 & 0 & 1 & 313.2488 & 17.7663 & 6803 & 7043 & 6450 & 1.017 \\
76 & 13.80 & 13.21 & 12.91 & 12.73 & 13.23 & 13.81 & 5 & 5 & 0.0164 & 28.9115 & 109.8058 & -32.6709 & 322.1726 & 19.1316 & 0 & 1 & 322.1729 & 19.1316 & 6903 & 7166 & 6566 & 1.017 \\
92 & 15.64 & 15.30 & 15.09 & 14.94 &  &  & 5 &  & 0.0208 & 13.1125 & 104.5148 & -47.9564 & 305.8667 & 16.2222 & 0 & 0 & 305.8667 & 16.2222 & 5376 & 5591 & 5015 & 1.013 \\
94 &  &  &  &  & 15.84 & 16.52 &  & 4 & 0.0336 & 80.6417 & 120.8356 & 17.9718 & 16.8900 & 17.6394 & 0 & 0 & 16.8900 & 17.6394 & 4098 & 4367 & 3995 & 1.011 \\
\dots \\
\hline
\multicolumn{20}{l}{$^\dag$ The complete version of this table is available online and also as a catalog within the Extragalactic Distance Database (\url{https://edd.ifa.hawaii.edu}).} \\
\multicolumn{20}{l}{$^a$ Average stellar mass from the $r$, $i$ and $z$ bands.} \\
\multicolumn{20}{l}{$^b$ Stellar mass from the WISE $W1$-band data.} \\
\multicolumn{20}{l}{$^c$ The mass of gas includes the contributions from \hi and Helium contents.} \\
\multicolumn{20}{l}{$^d$ Baryonic mass includes the stellar and gas contributions.} \\
\end{tabular}
\end{table}

\end{landscape}

\section{Summary}

Interstellar gas can be a significant component of the baryonic mass inventory of a spiral galaxy.  \citet{2000ApJ...533L..99M} have long been advocates of the BTFR as providing a more complete representation of galaxy mass and for the empirical justification that it provides a closely power law dependence with Hydrogen profile linewidths.

Our sample of $10^4$ galaxies gives good definition of the fundamental correlation.  As a tool for measuring galaxy distances, we had to overcome an evident bias.  Our sample is selected through \hi detections which are flux limited.  Photometry can be obtained for all candidates that pass the \hi bar.  The evidence for the bias is clear, as demonstrated in Figure~\ref{fig:delHI}. After accounting for this bias the distances that we measure are free of deleterious trends with redshift and pass the other tests of $\S\ref{sec:tests}$.

Our photometry is a mix of optical material from SDSS and infrared material from WISE, as discussed fully by \citet{2020ApJ...902..145K}.  The L-band infrared photometry is most cleanly linked to stellar mass. We derived linkages to stellar mass for the optical photometry that involve color terms.  There is good consistency between the optical and infrared material.

The optical and infrared observations that we use have different sky coverage.  The SDSS optical coverage is confined to the celestial north while the WISE infrared coverage is all-sky.  While the WISE photometry could be obtained for all our targets, in practice the coverage of objects with SDSS material is only partial.  The Cepheid and TRGB absolute calibrations result in optical and infrared BTFR that are slightly ($0.8\%$) different.  While this difference could be entirely statistical, it could arise from the difference in sky-coverage and differential departures from Hubble flow.  Since the goal of the program is to map non-Hubble motions, the separate optical and infrared calibrations are retained.  Of 10153 cataloged galaxies, 9984 are accepted with flag=1. Of these, 2095 galaxies have both optical and infrared measures, 4855 have only optical, and 3034 have only infrared.  A straight average of the SDSS and WISE distance moduli are taken if both are available. Observed rms scatter in moduli is about $0.45$ mag corresponding to $\sim20\%$ in distance.

The separate zero point fits to the BTFR give values of the Hubble Constant of $75.6\pm2.3$ and $74.1\pm2.8$~\kmsMpc\ for the infrared and optical cases, respectively.  An average of the optical and infrared distances gives $H_0=75.5\pm2.5$~\kmsMpc.
The uncertainties are dominated by the uncertainties of the zero point calibrations.  The optical and infrared field galaxy samples are 80\% independent and the optical and infrared calibration samples are 40\% independent.  This estimate of the Hubble Constant is provisional in the context of the {\it Cosmicflows-4} program.  It remains for these BTFR distances to be integrated with the seven other methodologies of the program, yet to be reported.

\section*{Acknowledgements}

We thank our anonymous referee for the constructive comments that certainly improved this presentation.
Special thanks to Cullan Howlett and Khaled Said for pointing out the trends in galaxy distances as functions of apparent optical and infrared magnitudes in \citet{2020ApJ...902..145K}, the impetus for the present study.  
Support for the {\it Cosmicflows-4} program has been provided by the NASA Astrophysics Data Analysis Program through grant number 88NSSC18K0424.

\section*{Data Availability}

We presented all data underlying this article in Table \ref{tab:catalog}. The complete version of this table would be available within the public domain of Extragalactic Distance Database (EDD: \url{https://edd.ifa.hawaii.edu/dfirst.php}) under the title of {\tt "CF4 BTF-distances"}. The corresponding photometry catalog is provided in table {\tt "CF4 Initial Candidates"} on EDD. The underlying \hi data are available in multiple tables on EDD within the section {\tt "HI Linewidths"}. For further details on how the photometry and \hi data were assembled, please refer to Section 2 of \citet{2020ApJ...902..145K}.

\bibliography{paper}
\bibliographystyle{mnras}

\bsp	
\label{lastpage}

\end{document}